\documentclass[3p]{elsarticle}

\usepackage{lineno}

\usepackage{graphicx}
\usepackage{amsmath,pifont,amssymb}

\usepackage{bm}      
\usepackage{color}

\usepackage{hyperref}
 
 \usepackage{subfig}
 \usepackage{caption}
 \usepackage[toc,page]{appendix} 
 
\newsavebox{\astrutbox}
\sbox{\astrutbox}{\rule[-5pt]{0pt}{20pt}}

\def\DEL#1{{\textcolor{green}{#1}}}        

\def\squarebox#1{\hbox to #1{\hfill\vbox to #1{\vfill}}}

\newcommand{\defin}{\stackrel{\scriptscriptstyle\triangle}{=}}

\newcommand{\bu}{\boldsymbol{u}}

\newcommand{\B}{\boldsymbol{B}}

\newcommand{\car}{1\!\mathsf{I}}

\newcommand{\transp}{^{\scriptscriptstyle T}}
\newcommand{\Matsqrt}{^{\frac{1}{2}}}
\newcommand{\MatsqrtT}{^{\frac{1}{2}{\scriptscriptstyle T}}}
\newcommand{\invMatsqrt}{^{-\frac{1}{2}}}

\newcommand{\Hop}{\mathbb{H}}
\newcommand{\Mop}{\mathbb{M}}
\newcommand{\Exp}{\mathbb{E}}
\newcommand{\Id}{\mathbb{I}}

\newcommand{\Hess}{\mathcal H}

\newcommand{\adj}{\lambda}
\newcommand{\mdl}{\mathbb{M}}
\newcommand{\inc}{{X}}
\newcommand{\incr}{{\delta\! X}}
\newcommand{\obse}{{\cal Y}}
\newcommand{\mobs}{\mathbb{H}}

\newcommand{\x}{{x}}
\newcommand{\X}{{X}}
\newcommand{\STinc}{{\cal V}}
\newcommand{\STobs}{{\cal O}}

\newcommand{\STnoiseo}{{\cal O}}
\newcommand{\hil}{{\cal V}}

\newcommand{\adjH}{\left(\partial_{\inc}\mobs \right)^{*}}

\newcommand{\adjM}{\left(\partial_{\inc}\mdl \right)^{*}}

\newcommand{\tanH}{\partial_{\inc}\!\mobs\;}
\newcommand{\tanphi}{\partial_{\inc}\varphi\;}
\newcommand{\tanM}{\partial_{\inc}\mdl}

\newcommand{\noise}{{\eta}}

\modulolinenumbers[5]
\journal{Journal of Computational Physics}

\begin{document}
\begin{frontmatter}

\title{Evaluation of an ensemble-based incremental variational data assimilation}

\author[Inria]{Yin Yang \corref{cor1}} 
\author[Inria]{Cordelia Robinson \corref{cor1}}
\author[Irstea,UEB]{Dominique Heitz \corref{cor2}}
\author[Inria]{Etienne M\'emin \corref{cor1}}

\address[Inria]{Inria, Campus universitaire de Beaulieu, 35042 Rennes Cedex, France, ({\tt name.surname@inria.fr})}
\address[Irstea]{Irstea, UR TERE, F-35044 Rennes, France, ({\tt name.surname@irstea.fr})}
\address[UEB]{Universit\'e europ\'eenne Bretagne, Rennes, France.}


\begin{abstract}
In this work, we aim at studying  ensemble based  optimal control strategies for data assimilation.  Such formulation nicely combines the ingredients of ensemble Kalman filters and variational data assimilation (4DVar). In the same way as variational assimilation schemes, it is formulated as the minimization of an objective function, but similarly to ensemble filter, it introduces in its objective function an empirical ensemble-based background-error covariance and works in an off-line smoothing mode rather than sequentially like sequential filters.
These techniques have the great advantage to  avoid the introduction of tangent linear and adjoint models, which are necessary for standard incremental variational techniques. They also allow handling a time varying background covariance matrix representing the error evolution between the estimated solution and a background solution. As this background error covariance matrix -- of reduced rank in practice -- plays a key role in the variational process, our study particularly focuses on the generation of the analysis ensemble state with localization techniques. Besides, to clarify well the differences between the different methods and to highlight the potential pitfall and advantages of the different methods, we present key theoretical properties associated to different choices involved in their setup. 
We compared experimentally the performances of several variations of an ensemble technique of interest with an incremental 4DVar method.  The comparisons have been leaded on the basis of a Shallow Water model and have been carried out both with synthetic data and  through a close experimental setup. The cases where the system's components are either fully observed  or only partially have been in particular addressed.
\end{abstract}


\end{frontmatter}
\linenumbers
\section{Introduction}
Data assimilation techniques aim at recovering a system state variables trajectory along time from partially observed noisy measurements of the system. These procedures, which couple dynamics and noisy measurements of the system, fulfill indeed a twofold objective. On one hand, they provide a denoising -- or reconstruction -- procedure of the data through a given model framework and on the other hand, they provide estimation procedures for unknown parameters of the dynamics. A multitude of unknown parameters may be considered. The most usual ones concern the initial condition, some parametric forcing, the inlet or boundary conditions or functions modeling the errors on the model. Among all these unknowns, the knowledge of an accurate initial condition is for instance crucial in geophysics for forecasting applications. For several  reasons, the initial condition cannot be fixed in a naive way directly from the data. First of all, the measurements are most of the time sparse and noisy. They are in addition local and do not live on the same grid as the model. Spatial smoothing, referred as Kriging methods, allows the reconstruction of a dense measurement field on a targeted grid \cite{Steindl01}. However, the observations may reflect physical phenomenon that are not encoded in the model, or incorporate noise and unphysical abnormal data caused by measurement device failures. A direct integration of such initial condition may lead to a fast divergence \cite{Artana12,Gronskis_etal_2013}. Consequently, overstating somehow the case, we can say that the data alone, without a model, do not allow characterizing a situation, whereas, a model without any data, provide no information on the actual system. Thus, the best answer lies in the middle way: in between the data and a model. 
 
Several data assimilation procedures have been proposed in the literature. They can be divided in two main families. The first one, ensues from stochastic filtering principles. They aim at recovering the whole conditional distribution of the state variables trajectory, given a series of measurements along time. When only past data are considered, the procedure is called a filter, whereas it is referred as a smoother when past and future measurements -- with respect to the current state variable -- are included. In geophysics, those techniques rely on Monte Carlo approximation of the Bayesian filtering equation or on sample approximation of the linear Gaussian filtering case. The former is named particle filters in the literature \cite{Gordon93} and the latter is referred as ensemble Kalman filters (EnKF) \cite{Evensen94}. Both techniques face a so called curse of dimensionality, which results from the sampling of a gigantic state space (often $\propto 10^{8}$ to $10^9 $) from only very few samples (at most $100$). Ignoring the stochastic modeling issue, this family of data assimilation procedures can be almost directly  implemented from any given numerical dynamics. The second family of data assimilation strategy is formulated as an optimal control problem \cite{Ledimet86,Lions71}. Here, the solution is expressed as the trajectory starting in the vicinity of a background solution and leading to the lowest data discrepancy. In the simplest case, it is an initial value problem. Formally, it is formulated as the minimization of an objective function. Due to the faced dimension, optimization based on the direct functional gradient computation cannot be implemented. An efficient and elegant optimization procedure relying on tangent linear expressions of the dynamics and data discrepancy operator has been proposed for that purpose. This scheme combined with nonlinear least squares strategies constitutes the principal ingredient of variational data assimilation procedure working routinely in operational centers \cite{Courtier94}. These techniques have proven to be efficient for numerous applications. Nevertheless, the derivation of an adjoint dynamics model may turn out to be an extremely tedious task, which, even with the help of automatic differentiation tools, requires skills and manpower. The other drawback lies in the discrepancy norm used between the unknown initial condition and the known background solution. It usually involves a stationary background covariance matrix. This stationary assumption is a crude hypothesis that is not considered for ensemble filters. This point and the fact that no adjoint model is required  constitutes two very attractive aspects of ensemble methods.    

Recently, several schemes aiming at coupling the advantages of ensemble methods and variational assimilation strategies have been proposed.
In the continuity of the ideas proposed for fixed time data assimilation cost function ({Hessian-incr}3DVar) \cite{Hamill00} and extended later to temporal variational data assimilation \cite{Lorenc03}, several authors have proposed methods that express explicitly the solution as a linear combination of the square root of the ensemble-based covariance\cite{Buehner05,Buehner10fq,Buehner10jr,Fertig07,Hunt04,Zhang09,Zupansky05,Clayton12,Fairbairn13}. The optimization is often led relying on ensemble Kalman filter update and eventually coupled with an adjoint scheme \cite{Buehner10fq,Buehner10jr,Zhang09,Clayton12}. Another notable point of view proposed by \cite{Desroziers12} tries to obtain 4D analysis based on 4D covariance matrix derived from ensemble.

In this work we chose to assess a method closely related to a strategy proposed by \cite{Buehner05} and \cite{Liu08}, which we referred as En4DVar in the following. This technique introduces in its objective function an empirical ensemble-based background-error covariance which avoids the use of tangent linear and adjoint model. The associated optimization is conducted  as a gradient descent procedure and does rely on iterative ensemble filtering updates exploiting equivalences between Kalman smoothers and the {\em a posteriori} energy minimization established only in the linear case. 

Beyond this practical implementation aspect, it is nevertheless difficult to state precisely the limitations of such a technique and to assess its comparative performances with respect to the standard 4DVar implementation. As a mater of fact, instead of considering the whole state space, ensemble 4D-Var methods provide, on one hand, solutions defined on a subspace spanned by the ensemble members, and, on the other hand, they introduce a flow dependent time varying covariance instead of a static full-rank background  covariance. As the background error covariance matrix plays a key role in the variational process, our study particularly focuses on the generation of the analysis ensemble state with localization techniques. An experimental evaluation between an incremental 4DVar and an ensemble based 4DVar will be carried out in this study. This evaluation is performed using a Shallow Water model.

This paper is organized as follow. In the first section, we will recall briefly the principles of variational assimilation techniques. Implementation aspects will also be described. In the second section we will focus on the description of an ensemble formulation of this variational assimilation technique. The last section will finally describe our experimental setup and the results we obtained. 

\section{Variational data assimilation}
This section aims at describing briefly the general principles of variational assimilation techniques. It will also be the occasion to fix notations and the vocabulary used in this paper. As recalled in the introduction, data assimilation techniques consist in estimating the evolution of a given ensemble of a system's state variables, denoted $X$, by associating measurements, $\obse$, of this system with a dynamic model describing the system's evolution. The state variables and the measurements are both assumed to live in two Hilbert spaces identified to their dual, noted respectively ${\cal U}$ and ${\cal V}$.  This can be summarized with the following equations: 
	\begin{align}
		&\partial_t X(t,x) + \mathbb{M}(X(t,x)) = 0, \label{Ass-syst-dyn}\\
		&X(t_0,x) = X^b_0(x) + \eta(x), \label{Ass-syst-data}\\
		&\obse(t,x) = \mathbb{H}(X(t,x)) + \epsilon(t,x) \label{Ass-syst-init}.
	\end{align}	
The first equation is the evolution model. It involves a  differential operator $\mathbb{M}$ -- usually nonlinear. 
A complete integration of this model on the assimilation window, starting from time $t_0$ to time $t_f$, provides a trajectory of the state variables. The dynamical model may be assumed to be completely known or to depend on unknown parameters.  An error function on the right-hand side of Eq.\eqref{Ass-syst-dyn} can be for instance possibly added to model uncertainty or errors of this dynamics.  The second equation defines the initial state, $X(t_0,x)$, assumed to be known up to a zero Gaussian noise, $\eta$, of covariance $B$. The integration of the background initial guess, $X^b_0$, is called the background trajectory, $X^b(t,x)$. It is usually fixed from a previous assimilation result on an anterior temporal window. The last equation links the observations, $\obse$, and the state variable $X$ through an observation operator $\mathbb{H}$. In this study, for simplification purpose, we will consider a linear operator set to  the identity or to an incomplete identity when only a part of the state is observable. The discrepancy error, $\epsilon$,  between the measurements and the state  is assumed to be a zero mean Gaussian random field with a covariance tensor $R$. This noise is assumed to be uncorrelated from the state variable. 

A standard variational data assimilation problem associated with system Eq.\eqref{Ass-syst-dyn} \eqref{Ass-syst-data} \eqref{Ass-syst-init} is formulated as the minimization of the following objective function with respect to the unknown initial condition discrepancy:
\begin{equation}
	J(\eta(x)) = \\\frac{1}{2} \displaystyle \| X^b_0(x) - X(t_0,x) \|^2_{B}  + \frac{1}{2} \displaystyle \int_{t_0}^{t_f} \|  \mathbb{H}(X(t,x)) - \obse(t,x) \|^2_{R} dt.
	\label{Jnoinc}
\end{equation} 
This objective function involves the $L_2$ norm with respect to the inverse covariance tensor 
\[\|f\|^2_A= \int_{\Omega} f(x) A^{-1}(x,y) f(y) dxdy.\] 
The associated minimization problem is referred to the literature as the strong constraint variational assimilation formulation. This constitutes an optimal control problem where one seeks for the value of  the initial condition discrepancy, $\eta$, that yields the lowest error between the measurements and the state variable trajectory.
Let us note that such an energy function can be interpreted as the log likelihood function associated to the {\em a posteriori} distribution of the state given the past history of measurements and the background. Denoting $\varphi_t(X_0)=X_t= X_0 + \int_0^t \Mop(X(s)) ds$ as the flow map (viewed here as a function of a random initial condition), and assuming it is a diffeomorphism (e.g. a differentiable map whose inverse exists and is differentiable as well), we get 
\begin{align}
p(X_t | Y_{t_f}, \ldots, Y_{t_0}, X_{t_0}^b )  \propto& p(Y_{t_f}, \ldots, Y_{t_0}| \varphi_t(X_{t_0}), X^b_0 ) p(\varphi_{t}(X_{t_0})| X^b_{0} ), \nonumber\\
\propto& \prod\limits_{t_i=0}^{t_f} p(Y_{t_i}|\varphi_{t_i}(X_{t_0}))p(\varphi_{t}(X_{t_0})| X^b_{0} ), \nonumber\\
\propto& \prod\limits_{t_i=0}^{t_f} p(Y_{t_i}|\varphi_{t_i}(X_{t_0}))p( \varphi_{t}^{-1}(X_{t}) | X^b_0)|\rm{det}\partial_X\varphi_{t}^{-1}(X_t)|, \nonumber\\
\propto& \prod\limits_{t_i=0}^{t_f} p(Y_{t_i}|\varphi_{t_i}(X_{t_0}))p( X_{t_0} | X^b_0)|\rm{det}\partial_X\varphi_{t}^{-1} (X_t)|,
\label{AP-pdf}
\end{align}
where $ \partial_X\Mop (X)$ denotes the tangent linear operator of $\Mop$, defined as
\begin{equation}
\lim_{\beta\rightarrow0} \frac{\Mop(X'+\beta dX)-\Mop(X')}{\beta} =
\partial_{X} \Mop(X')dX .
\end{equation}
 The argument indicates, when clarity is needed, the point at which this derivative is computed. It is straightforward to check that the linear tangent of a linear operator is the operator itself. 
We assumed here that observations at a given time depend only on the state at the same time, and that they are conditionally independent with respect to the state variables. Now, replacing the probability distribution by their expressions Eqs \eqref{Ass-syst-data} and \eqref{Ass-syst-init}, we get  the objective function Eq.\eqref{Jnoinc} up to a time dependent factor. If the flow map is volume preserving and invertible, we have $|\rm{det}\partial_X\varphi_{t}^{-1} (X_t) |=|\rm{det}\partial_X\varphi_{t}(X_{t_0}) |=1 $ and we get exactly the sought energy function. In that case, the maximum {\em a posteriori} estimate hence corresponds to the objective function minima. In the Gaussian case, the maximum {\em a posteriori} estimate and the conditional mean with respect to the whole measurements\footnote{It is a well-known fact that this estimate constitutes the minimum variance estimate\cite{Anderson79}.} trajectory are identical. Let us note, nevertheless, that the {\em a posteriori} pdf is Gaussian only if the dynamical model and the observation operator are both linear. We also point out that even the tangent linear expression of the dynamical models, as defined in geophysical applications, are not invertible in general. For a time and space discrete approximation of the tangent linear dynamical operator ({\em i.e.} a matrix), a pseudo inverse may be defined from a Singular Value Decomposition (SVD). Noting $\partial_\inc \varphi_t (X'_0)$ this matrix operator, which depends on time and on a given initial condition, we can write
\begin{equation}
\partial_\inc \varphi_{t}(X'_0) = U(t,X'_0) \Sigma(t,X'_0) V\transp(t,X'_0),
\end{equation}
where $\Sigma(t,X(0))=\rm{diag}(\sigma_1, \cdots,\sigma_p,0\cdots,0)$ is a  real diagonal matrix gathering the square-root of the eigenvalues of the matrix $\partial_\inc\varphi\transp_t\partial_\inc\varphi_t$, and both U and V are orthonormal matrices, corresponding respectively to eigenvectors of matrix $\partial_\inc\varphi\partial_\inc\varphi\transp$  and $\partial_\inc\varphi\transp\partial_\inc\varphi$ respectively. The three matrices are $n\times n$ and may depend on time and on the initial condition.  The pseudo inverse is 
\begin{equation}
\partial_\inc\varphi_{t}^{-1}(X_0) = V(t,X_0) \Sigma^{-1}(t,X_0) U\transp(t,X_0),
\end{equation}
where $\Sigma^{-1}(t,X'_0)= \rm{diag}(1/\sigma_1, \cdots,1/\sigma_p,0\cdots,0)$ and for divergence free volume preserving dynamics we get 
$|\rm{det}\varphi_{t}^{-1} (X_t)|  = |\prod\limits_{i=1}^p \sigma_i^{-1}|= 1$ and the logarithm of the posterior  corresponds to objective function Eq.\eqref{Jnoinc}.

\subsection{Properties of variational data assimilation}
We recall in this subsection some well-known properties of the standard variational functional that will be useful in the following (see \cite{Li-Navon-01,Lorenc86,Rabier92}). A (local) minimizer of functional Eq.\eqref{Jnoinc} is provided for $\partial_\eta J(\hat{\eta})=0$. The gradient with respect to the initial condition is given by (see appendix A):
\[
\partial_{\eta} J (\eta)= B^{-1}(X_0-X_0^b) + \int_{t_0}^{t_f}  \partial_{X} \varphi^*_t \partial_X\Hop^* R^{-1}(\Hop X(t) -\obse(t))  dt,
\] 
and the  minimizer reads 
 \begin{equation}
 \hat{X}_0 = X_0^b + \Hess^{-1} \partial_\eta J ( X_0^b),
 \label{update}
\end{equation}
where $\Hess$ is the Hessian matrix gathering the cost function second derivatives:
\begin{equation}
\Hess = B^{-1} + \int_{t_0}^{t_f}  \partial_X\varphi^*_t \partial_X\Hop^* R^{-1} \partial_X\Hop \partial_x \varphi_t dt.
\end{equation}

When the measurement and the dynamics operators are both linear, the objective function is convex, and this estimate, if it is accurately computed, corresponds to the unique global minimum. Denoting the true state $X_0^t$, we have $\nabla  J (\hat{X}_0- X_0^t + X_0^t)=0 \Leftrightarrow$
\begin{multline} 
( B^{-1} + \int_{t_0}^{t_f}  \partial_{X} \varphi^*_t \partial_X\Hop^* R^{-1} \partial_X\Hop \partial_{X} \varphi_t dt)(\hat{X}_0- X_0^t) \approx\\B^{-1}(X_0^b- X_0^t) + \int_{t_0}^{t_f}  \partial_{X} \varphi^*_t (X_0)\partial_X\Hop^* R^{-1}(\obse(t) - \Hop (\varphi_t (X^t_0)))  dt,
\end{multline}
where a linearization  around the true state, $X^t$, of the cost function gradient at the minimum point has been performed. A strict equality only applies for linear operator. 
Multiplying the left and right terms by their transpose expression, taking the expectation  and assuming the background errors $X^b_0-X_0^t$ and the innovation $\obse(t) - X^t(t)$
are uncorrelated, we get, the classical result:
\begin{align}
P_0\approx \left ( B^{-1} + \int_{t_0}^{t_f}  \partial_{X} \varphi^*_t \partial_X\Hop^* R^{-1} \partial_X\Hop \partial_{X} \varphi_t dt \right ) ^{-1} = \Hess^{-1}, 
\label{inv-Hess}
\end{align}
which states that the initial error covariance matrix $P_0=\Exp ((X_0^b- X_0^t)(X_0^b- X_0^t)\transp)$ is given by the inverse Hessian matrix of the functional. This matrix is usually called the analysis covariance matrix, and is specifically denoted by a superscript "{\em \!a}". From distribution Eq.\eqref{AP-pdf}, for a volume preserving transformation, we observe immediately that an optimal initial value supplies an optimal trajectory that maximizes the {\em posterior} distribution.
Thus, at a given time, the analysis error covariance reads (assuming without loss of generality a zero mean): 
\begin{equation}
P^a_t=\int (\varphi_t(X_0^b- X_0^t)\varphi\transp_t(X_0^b- X_0^t)) p(X_t | Y_{t_f}, \ldots, Y_{t_0}, X_{t_0}^b )  dX_t.
\end{equation}
If the dynamical model is linear, $\varphi_t(X_0)= \Phi_t X_0$, we then get:
 \begin{equation}
P^a_t=\Phi_t P_0^a \Phi_t\transp,
\end{equation}
which describes the way the errors are propagated along time in  variationnal data assimilation. This formula also unveils a recursion, involving a forecast state $\bar{X}^f_{t}= \Mop \hat{X}_{t-dt}$ and a forecast covariance matrix $P_t^f=\Mop P_{t-dt}^a \Mop\transp $. In the linear case, with an additive  stationary noise, the update expression  can be written as:
\begin{align}
\hat{X}_t =& \bar{X}^f_{t} + K_t(\obse_t - \Hop X_t^f),\\
K_t=& ({P_t^{f}}^{-1} + \Hop\transp R^{-1}\Hop)^{-1},\\
P^a_t=& (\Id - K\Hop) P^f_t,
\label{K-eq}
\end{align}
which is the well known recursive Kalman filter formulation (associated to a noise free dynamics). A more usual  Kalman gain matrix formulation can be obtained through the Sherman-Morrison-Woodbury formula: 
\begin{equation}
K_t= P^f_t \Hop\transp ( R + \Hop P^f_t \Hop\transp)^{-1}.
\end{equation}
This expression of the Kalman gain is computationally advantageous as the inversion involved is performed in the observation space, which is usually of lower dimension than the state space. A recursive Kalman filter with a deterministic perfect dynamics is thus equivalent to the previous variational assimilation strategy only at the final time of the assimilation window. Within the assimilation interval, the variational assimilation and the Kalman filter differ. The former corresponds to a smoothing procedure as the whole set of measurements are taken into account, whereas the latter relies only on the past data. 

In high dimensional application, a brute force application of these equations cannot be implemented. Efficient ensemble techniques have been devised specifically for that purpose following on from the work initiated by Evensen \cite{Evensen94}.   They are mainly defined through empirical expression of the forecast mean and covariance matrix and an efficient computation  through incomplete SVD of the update equation. Two main kinds of methods have been proposed for that purpose. The first one relies on a direct Monte Carlo approach, which introduces measurement noise samples \cite{Burgers98,Evensen96, Houtekamer98}. The second one, corresponds to square-root filter \cite{Anderson03, Bishop01, Ott04, Tippett03, Whitaker02}. Those latter schemes avoid  sampling issues associated to small size ensembles by confining the analysis in the space spanned by the forecast ensemble centered perturbations. This is operated through the introduction of the square-root of the analysis error covariance matrix. One example of such a procedure is the Ensemble Transform Kalman Filter (ETKF), proposed initially by \cite{Bishop01}, which consists to assume that the estimate (i.e. the posterior mean) lives in the subspace generated by $p$ samples:  $\hat{X}_t= \bar{X}^f_t + X^f_t w$,  where $X^f_t$ denotes the anomaly matrix $X^f_t= [X^{f,(1)}_t - \bar{X}^f_t, \cdots, X^{f,(p)}_t - \bar{X}^f_t]$ and the vector, $w$ corresponds to the coordinates in the subspace spanned by the  sampled ensemble members. In the ETKF technique  $w$ is fixed in order to respect the Kalman equations Eq.\eqref{K-eq}:
\begin{align}
\hat{X}_t =& \bar{X}^f_{t} +X^f_t w,\\
w=  &A_t A_t\transp (\Hop X^f_t)\transp R^{-1}  (\obse_t - \Hop \bar X_t^f),\\
A_t = &[(p-1)(\Id_p + (\Hop X^f_t)\transp R^{-1}\Hop X^f_t)]\invMatsqrt,\\
P^a_t=& X^f_t A_t A_t\transp {X^{f}_{t}}\transp.
\label{ETKF-eq}
\end{align}
We remark here, that the observation operator is possibly nonlinear. Note also that from an implementation point of view, a singular value decomposition of matrix $(\Hop X^f_t)\transp R^{-1}\Hop X^f_t$ is usually performed. In addition, the square root matrix $A$ is not unique and can be defined up to an orthogonal matrix. Judicious choices allow preserving in the analysis matrix the zero mean of the forecast anomaly matrix ({\em i.e.} $X^f_t \car =0 \Rightarrow X^f_t A_k \car=0$) \cite{Wang04,Sakov07}. 

Let us note that other ensemble filters, defined from the particle filtering concept \cite{Gordon93},  that use these recursive equations to build efficient procedures have been proposed in the literature \cite{Beyou13a, Beyou13b,Papadakis10,VanLeeuwen-10}.

\subsection{Functional minimization}
Due to the dimension of the state space, the minimization of the function (\ref{Jnoinc}) requires implementing an iterative optimization strategy. The most efficient  optimization procedure needs to evaluate the functional gradient at several points. A direct numerical
evaluation of the data assimilation objective function gradient  is unfortunately computationally unfeasible, as it requires the computation of perturbations of the state variables along all the components of the control variables ($\eta$) (that is to say to integrate the dynamical
model for all perturbed components of the initial condition, which is obviously not possible in practice). An elegant solution to this problem consists in relying on an adjoint formulation \cite{Ledimet86,Lions71}. Within this formalism, the gradient functional is obtained
by a forward integration of the dynamical system followed by a backward integration of an adjoint variable, $\lambda$, which is driven by dynamics defined from the adjoint of the tangent linear dynamical operator, $\partial_X \Mop$, and the tangent linear observation operator, $\partial_X \Hop$. This adjoint backward dynamics, defined as:
\begin{equation}
\left\{
\begin{array}{l}
-\partial_t \adj(t)+\adjM \adj(t) = \adjH R^{-1}(\obse-\mobs(\inc(t)))\\
\;\adj(t_f) = 0,
\end{array}
\right.
\label{dyn-adj}
\end{equation}
\noindent
provides the gradient functional  at the initial time (see appendix for a complete description):
\begin{equation}
\partial_\eta J (X_0) = -\lambda (t_0) + B^{-1} (X_0- X^b_0).
\label{update-adj}
\end{equation}
At this point, an immediate quasi Newton optimization procedure emerges, where the estimate  at iteration $n+1$ is updated as:
\begin{equation}
\X^{n+1}_0 = X^{n}_0 - \alpha_n \tilde \Hess_{X^n}^{-1} \partial_{\eta} J (X^n_0),
\end{equation}
where  $\tilde \Hess_{X^n}^{-1}$  denotes an approximation of the Hessian inverse computed from the functional gradient with respect to $X^n$ ; the constant $\alpha_n$ is chosen so that the Wolfe conditions are respected. The adjoint variable is accessible
through a forward integration of the state dynamics Eq.\eqref{Ass-syst-dyn} and a backward
integration of the adjoint variable dynamics Eq.\eqref{dyn-adj}.  This constitutes the standard 4Dvar data assimilation procedure. An algorithmic synopsis of  the 4DVar is described in algorithm Fig. \ref{algo-assim}. 

\begin{figure}[h!]
\begin{center}
\fbox{\begin{minipage}[t]{0.8\linewidth}
\begin{enumerate}
\item Set an initial condition: $\inc(t_0) = \inc_0$
\item From  $\inc(t_0)$, compute $\inc(t)$ with the forward
  integration  of relation (\ref{Ass-syst-dyn})
\item Compute the adjoint variable $\adj(t)$ with the backward
  integration  of relation (\ref{dyn-adj})
\item Update the initial value $\inc(t_0)$ 
  with (\ref{update-adj})
\item Loop to step $2$ until convergence
\end{enumerate}
\end{minipage}}
\end{center}
\caption{Schematic representation of the variational data-assimilation algorithm}
\label{algo-assim}
\end{figure}

\subsection{Incremental variational assimilation}
When the operators involved are nonlinear, the variational assimilation  procedure can be improve by introducing a nonlinear least squares procedure in the same spirit as a Gauss-Newton incremental strategy\cite{Courtier94}. This optimization strategy consists in performing a linearization of the dynamics around a current trajectory and operating the optimization with respect to an incremental solution. Instead of correcting directly the initial state $X(t_0)$ as in the previous subsection, the incremental approach consists in correcting an increment, $\delta X_0$, that evolves according to the tangent dynamics equation computed around a given state $X$:
\begin{equation}
	\left \lbrace
		\begin{array}{l}
			\partial_t \incr(x,t) + \partial_{X} \mathbb{M} (\inc)\incr (x,t) = 0 \\
			\incr(x,0) = \inc^b_0 - \inc_0 + \eta\\
		\end{array}
	\right.
	\label{incremental}
\end{equation}

Different variants of the incremental technique can be defined depending on the state around which the linearization is applied. It is possible for instance to choose to linearize around the same initial background at each step and to add a correction term to the innovation term \cite{Weaver03}. This technique comes to correct the background at each step. Another solution consists in updating the linearization state with the increment computed at each step \cite{Haben11}. In this case the innovation term is unchanged but a correction term is added to the energy function's background error term. This correction term constraints the whole initial condition, $X_0 + \incr_0(x)$,  to stick to the initial background. It can be shown that these two choices are indeed equivalent. In our case, we employ a slightly different approach. As a matter of fact, in this study we will focus on cases for which only a background of bad quality  is available. Such case typically corresponds to an image based assimilation problem in which the background is given by a noisy image with possibly large areas of missing data. 
Hence, we do not want to stick to it too much. At each step $X^b$ will be set to $X$, and thus $\incr(x,0)= \eta$. Let us outline, that in the ensemble approach we will present later, we will not only update the background state around which the 4DVar functional  is linearized but also the associated background error covariance matrix.  

The incremental  cost function we are considering  is consequently defined as:
\begin{align}
	J(\incr_0) =\frac{1}{2}  \|\incr_0(x) \|^2_{B}  + \frac{1}{2} \displaystyle \int_{t_0}^{t_f} \displaystyle  \| \tanH \incr (t,x)- D(t,x) \|^2_{R} dt,
		\label{Jinc}
\end{align}
where the innovation vector $D(t,x)$ is defined as:
\begin{equation}
D(t,x)=\obse(t,x) - \mathbb{H}(X (t,x)).
\end{equation}
We note that this incremental functional corresponds to a linearization around a given background  solution of the innovation term. It  is now quadratic and thus, has a unique minimum. This leads naturally to an optimization strategy that proceeds into two interleaved loops. An external loop corresponds to the evolution of the background solution through the nonlinear dynamics, while an internal loop computes for this solution an optimal increment driven by the tangent linear dynamics. The computed initial increment allows us to update the current background initial solution. In practice, due to the high computational cost associated with the  dynamical models involved in geophysical applications, only a low number of external iteration are performed. The adjoint system associated to this functional is directly inferred from the previous one and  reads:
\begin{equation}
\left\{
\begin{array}{l}
-\partial_t \adj(t)+\adjM \adj(t) = \adjH R^{-1}(D(t,x) - \tanH \partial_{X} \varphi_t(X_0) \delta\inc_0)\\
\;\adj(t_f) = 0,
\end{array}
\right.
\label{dyn-adj-incr}
\end{equation}
with the gradient function at the initial time (which can readily be deduced  from the results in the appendix):
\begin{equation}
\partial_{\eta} J (\incr_0) = -\lambda (t_0) + B^{-1} \delta X_0.
\label{update-adj-inc}
\end{equation}
This incremental scheme is summarized in Fig. \ref{algo-incr}. 
\begin{figure}[h!]
\begin{center}
\fbox{\begin{minipage}[t]{0.8\linewidth}
\begin{enumerate}
\item Set an initial state: $\inc(t_0)=\inc^b(t_0) $
\item From  $\inc(t_0)$, compute $\inc(t)$ with the forward
  integration  of the nonlinear dynamics (\ref{Ass-syst-dyn})
 \item Set the initial increment as: $\incr_0= 0$
 \item Compute the tangent linear dynamics (\ref{incremental})
\item Compute the adjoint variable $\adj(t)$ with the backward
  integration  of relation (\ref{dyn-adj-incr})
\item Update the initial value $\incr(t_0)$ 
  with (\ref{update-adj-inc})
\item Loop to step $4$ until convergence
\item Update the background: $X(t_0) +:=\delta \inc(t_0) $
\item Loop to step 2 for a specified number of iterations
\end{enumerate}
\end{minipage}}
\end{center}
\caption{Schematic representation of the incremental variational data-assimilation algorithm}
\label{algo-incr}
\end{figure}
The corresponding function (\ref{Jinc}) can be rewritten in a simpler  form as: 
\begin{equation}
J(\delta X_0) =\frac{1}{2} \Big<\delta X_0, {\cal H}\delta X_0\Big\rangle - \int_{t_0}^{t_f} \Big<\delta X_0,b\Big>  + C,
\label{Jincr}
\end{equation}
where the Hessian, the linear term and the constant are respectively defined as:
\begin{align}
&\Hess = B^{-1} + \int_{t_0}^{t_f}  \partial_X\varphi^*_t \partial_X\Hop^* R^{-1} \partial_X\Hop \partial_\inc\varphi_t dt, \label{Hessian-incr}\\
&b=  \int_{t_0}^{t_f}  \partial_{X} \varphi^*_t \partial_X\Hop^* R^{-1} D(t)  dt, \\
&C =  \int_{t_0}^{t_f}  D\transp(t)\partial_X\varphi^*_t \partial_X\Hop^* R^{-1} \partial_X\Hop \partial_\inc\varphi_t D(t)dt. 
\end{align}
We highlight that for a linear dynamics and linear observation operator, fixing the background covariance matrix to an empirical matrix $X_t^{f}{X_t^{f}}\transp$ with $X_t^{f}=\Phi_t X_0$ and where $X_0$ corresponds to an ensemble of initial conditions, we recover the Kalman filter equivalence at the end of the assimilation filter. Defining furthermore the Hessian (or the inverse of the error covariance) from the square root of Eq.\eqref{Hessian-incr} we get exactly the ETKF solution. This equivalence holds only for a fixed single time assimilation window (the 3DVar technique) or at the end of the assimilation window for a linear dynamics.  The ETKF technique may be useful to provide an approximate cost function's Hessian. 

The minimization of the cost function Eq.\eqref{Jincr} comes to solve a linear system $\Hess \incr_0 = b$. The possibly ill-conditioned nature of such a system depends on  the condition number of the Hessian matrix. As this matrix is here symmetric positive definite, 
the condition number is given by the ratio of the largest eigenvalue over the minimum eigenvalue.  The larger the condition number is, the more sensitive the system is to errors both in the $b$ vector and in the estimate.  
A known solution to improve  a badly-conditioned system, consists in solving an equivalent system with a lower condition number. Preconditioning constitutes a practical procedure to reach that goal. 

 \subsection{Variational assimilation Preconditioning}
 \label{subsec:Preconditioning}
Proceeding to a preconditioning with matrix $P$ of our variational incremental system consists in applying a change of variable with the matrix square-root of P:
\begin{equation}
\incr_t  = P\Matsqrt \delta Z_t ,
 \end{equation}
 such that the new Hessian matrix, $\widetilde{\Hess}= P\MatsqrtT \Hess P \Matsqrt$, possesses a lower condition number. For instance, we see immediately that if $P$ is set to the inverse of the Hessian matrix, then the original system is solved in a single step. This approach, which leads to the lowest unity condition number, is obviously unpractical since it requires solving the system. In variationnal assimilation, a common procedure consists to operate a preconditioning with the  background correlation matrix. This is called the control variable transform(CVT): 
 \begin{equation}
\incr_t = B\Matsqrt  \delta Z_t.
\label{CVT}
 \end{equation}
 Noting that the Hessian Eq.\eqref{Hessian-incr} has a first component that is the background information matrix, the background covariance matrix can be considered as a rough approximation of the Hessian inverse. To highlight its role on the Hessian conditioning, the condition number of the Hessian matrix can be computed in a simple case. Let us assume in the following that the measurements noise are independent and  spatially identically distributed with a constant covariance $R=\sigma \Id$ (and $\sigma$ constant). We  assume also that the measurements and the state variable live in the same space ($\Hop=\Id$). The Hessian condition number can be bounded in the matrix $2$-norm by:
 \begin{align}
 \|\Hess\|_2 &\leq \lambda_{\max} (B^{-1}) +  \sigma ^{-1}\int_{t_0}^{t_f}  \|C_t\|_2,\\
 &\leq  \lambda_{\max} (B^{-1}) + \sigma ^{-1}\max_{t\in [t_0 t_f]}(\lambda_{\max}(C_t))(t_f-t_0),
 \end{align}
 where $\lambda_{\max} (A)$ is the largest real  eigenvalue of a symmetric matrix $A$ and $C_t = \partial_\inc\varphi_t\transp \partial_\inc\varphi_t$. As matrix $C_t$ is in general rank deficient, its lower eigenvalue is zero and  we obtain immediately a lower bound of the Hessian 2-norm: 
 \begin{equation}
 \|\Hess\|_2\geq \lambda_{\min} (B^{-1}).
 \end{equation}
 These two bounds provide a bound on the Hessian condition number:
 \begin{align}
 \kappa(\Hess) &\leq \frac{\lambda_{\max}(B^{-1})}{\lambda_{\min}(B^{-1})}(1 + \frac{\max_{t\in [t_0 t_f]}(\lambda_{\max}(C_t))(t_f-t_0)}{\sigma\lambda_{\max}(B^{-1})}) \nonumber,\\
 \kappa(\Hess) &\leq \kappa(B)(1+ \sigma^{-1}\lambda_{\min}(B)\max_{t\in [t_0 t_f]}(\lambda_{\max}(C_t))(t_f-t_0)). 
 \end{align}
We observe that this bound depends directly on the background matrix. An ill-conditioned background matrix is likely to lead to an ill-conditioned Hessian. The Hessian conditioning bound depends also on the larger eigenvalue of matrix $C_t$, which can be related to the maximum finite time Lyapunov exponent on the flow domain and over the assimilation interval. Dynamics exhibiting high stretching integrated on large temporal window leads potentially to an ill-conditioned Hessian. Non noisy perfect measurements are also a potential source of bad conditioning. This situation can be well understood through pdf  Eq.\eqref{AP-pdf}. As a matter of  fact, considering a perfect measurement with  a variance tending to zero, gives a likelihood that tends to a Dirac  function. The log posterior pdf (\ref{AP-pdf}) is:
\begin{equation}
\int_{t_0}^{t_f} \delta(\obse_t - \varphi_t (\inc_{t_0}) )P(\varphi_t(\inc_{t_0})|\inc^b_{t_0}) dt. 
\end{equation}
 This pdf is maximal for the optimal solution and zero otherwise. If the measurements corresponds exactly to the model dynamics -- which is never the case in practice -- the optimal trajectory is given by $\varphi_t(\obse_{t_0})$. 
 However, tiny perturbation of the initial condition cancels the pdf and makes the estimation highly unstable. 
 
Considering now a  preconditioning  with the background covariance, we get a modified Hessian:
 \begin{equation}
\widetilde{\Hess} = \Id +   \int_{t_0}^{t_f}  B\MatsqrtT \partial_X\varphi^*_t \partial_X\Hop^* R^{-1} \partial_X\Hop \partial_\inc\varphi_t B\Matsqrt dt.
  \end{equation}
  With exactly the same assumption as previously, we obtain a  tighter bound for the modified Hessian:
  \begin{equation}
  \kappa(\widetilde{\Hess})\leq (1+ \sigma^{-1}\lambda_{\max}(B) \max_{t\in [t_0 t_f]}(\lambda_{\max}(C_t))(t_f-t_0)).
  \end{equation}
The background conditioning forms a potentially good candidate for the system preconditioning. We used here simplified assumptions for the observation operator and for the measurement noise. However, similar bounds can be established for a more general model assuming the observation space is of lower dimension than the state space \cite{Haben11}. As shown in this section, the background covariance matrix plays a central role. In the following, we describe a technique in which this covariance matrix is empirically determined. 
 
\section{Ensemble-based 4D Variational assimilation}

%
%
%
%
Applying the CVT \eqref{CVT} to the function \eqref{Jinc}, we get a modified objective function
\begin{equation}
J(\delta Z_0) =\frac{1}{2}  \| \delta Z_0  \|^2  + \frac{1}{2} \displaystyle \int_{t_0}^{t_f} \displaystyle  \| \tanH \tanphi_t  (\inc_0) B\Matsqrt\delta Z_0- D(t,x) \|^2_{R-1} dt.
\label{JincZ}
\end{equation}
This modified cost function removes $B^{-1}$ from the background term. This variable change can also be  viewed as a whitening filtering of the background error. Although better conditioned, the resulting system remains difficult  to solve and requires the use of the adjoint dynamics which  can be difficult to set up in practice.
Based on an empirical description of the background covariance matrix, \cite{Buehner05} and \cite{Liu08} proposed a technique that  precludes the use of the adjoint model to compute the gradient functional. We describe this technique hereafter. 

\subsection{Low rank approximation of the background error covariance matrix}
A low rank approximation of the background covariance matrix will allow us to formulate the variational ensemble technique. This low-rank approximation is introduced by the mean of initial state's samples. Denoting as $\langle f(t)\rangle =N^{-1} \sum_1^N f^{(i)}(t)$ an empirical ensemble mean of a quantity $f(t)$ through $N$ samples, the empirical background covariance matrix is:
\begin{equation}
B\approx \langle (\inc^b-\langle \inc^b \rangle)(\inc^b-\langle \inc^b\rangle)^T\rangle.
\end{equation}
Noting 
\begin{equation}
X'_{b}:=\frac{1}{\sqrt{N-1}}(\inc^{(1)b}_0-\langle \inc^b_0\rangle,\cdots,\inc^{(N)b}_0-\langle \inc_0^b\rangle ), \label{Sq_EnBack}
\end{equation}
the perturbation matrix gathering the $N$ zero mean centered background ensemble members as a low-dimensional approximation of the background matrix which can be compactly written as the following matrix form
\begin{equation}
B\approx \widetilde B =X'_b X'^{\scriptsize T}_b.
\end{equation}
This $n\times n$ matrix is at most of rank $N-1$ as the sum of its column vectors is null by construction.
Introducing the background covariance approximation in the preconditioned  cost function (\ref{JincZ}), we get
\begin{equation}
J(\delta z_0)= \frac{1}{2}  \| \delta Z_0 \|^2   + \frac{1}{2} \displaystyle \int_{t_0}^{t_f} \displaystyle  \| \tanH \tanphi_t  (\inc_0) X^{'}_b\delta Z_0- D(t,x) \|^2_{R} dt.
\label{JincZ-Bens} 
\end{equation}

Note that as in the standard incremental formulation all the non-linearity of the original observation operator $\Hop$ and of the model operator are maintained in the innovation process $D_t$.
 
\subsection{Functional minimization}
It is important to note that in the function (\ref{JincZ-Bens}), the term $\tanphi_t  (\inc_0) X^{'}_b$ corresponds to a forecast by the dynamical model of the centered square root background covariance matrix. This integration cannot be implemented on the full background covariance in practice. A direct minimization of this energy function would require the use of the adjoint minimization mechanisms. However, as we rely here on an empirical description of this matrix from a set of samples, we can observe that integrating these samples in time provides us immediately an empirical expression of a low-rank approximation of the  background covariance trajectory and of its square root. With this flow dependent background matrix, the functional is now 
\begin{align}
J(\delta Z_0) &= \frac{1}{2}  \| \delta Z_0  \|^2   + \frac{1}{2} \displaystyle \int_{t_0}^{t_f} \displaystyle  \| \tanH \widetilde B\Matsqrt_t\delta Z_0- D(t,x) \|^2_{R} dt,\label{JincZ-envar-1}\\
\widetilde B\Matsqrt_t &= \tanphi_t  (\inc_0) X^{'}_b.
\label{JincZ-envar-2} 
\end{align}
The gradient of this functional is given by:
\begin{equation}
\partial_ {\eta} J =  \delta Z_0 +\int_{t_0}^{t_f} \widetilde B\MatsqrtT_t \adjH \ R^{-1}(\tanH \widetilde B\Matsqrt_t\delta Z_0- D(t,x)) dt, \label{GincZ-envar}
\end{equation}
and the Hessian is:
\begin{equation}
\widetilde \Hess = \Id +  \int_{t_0}^{t_f} \widetilde B\MatsqrtT_t \adjH \ R^{-1}\tanH \widetilde B\Matsqrt_t dt.
\end{equation}
Once  the minimizer $\widehat{\delta Z}_0$ is estimated, the analysis increment is obtained by:
\begin{equation} 
\delta \inc_0 = \inc_0^b + \widetilde X'_b \widehat{\delta Z}_0.
\end{equation}
Let us emphasis that, as the covariance matrix $\widetilde B$ is at most of rank $N-1$, the control variable has at most $N-1$ non null component in the eigenspace. Compared to the full 4DVar approach, the control variable's degree of freedom is considerably lowered and the minimization computational complexity is significantly decreased. Indeed, this ensemble version has a lower computation cost if the ensemble forecasting step is distributed on a grid computing. A synopsis of the ensemble variational assimilation algorithm is provided in Fig. \ref{algo-ens4var}.

\begin{figure}[h!]
\begin{center}
\fbox{\begin{minipage}[t]{1\linewidth}
\centering{\em Outline of the ensemble incremental variational data-assimilation}

\begin{enumerate}
\item Set an initial state: $\inc(t_0)=\inc^b(t_0) $, initialize ensemble around this state
\item Parallelizing compute $\mathbf \inc(t)$: [$\inc(t), \inc_0^{(1)}, \cdots,  \inc_0^{(N)}$] from  $\inc(t_0)$ and each ensemble member $\inc_0^{(p)}$  with the forward integration of the nonlinear dynamics (\ref{Ass-syst-dyn})
\item Calculate the innovation vector $D(t,x)$
\item Derive the background perturbation matrix $X'_b$ from (\ref{Sq_EnBack}) and $\widetilde B\Matsqrt_t$ from (\ref{JincZ-envar-2})
 \item Initialized the initial increment: $\incr_0$
 \item Do an inverse control variable transformation $\delta Z_0= (X'_b)^{-1} \incr_0$ if necessary 
 \item Optimize $\delta Z_0$ in the inner loop, the cost function and the gradient are calculated based on (\ref{JincZ-envar-1}) and (\ref{GincZ-envar}) respectively
 \item Update control initial space $Z_0$ and calculate $\incr_0$  by transforming $\delta Z_0$ to the state space with  (\ref{CVT}) 
\item Update the initial condition: $X(t_0) +:= \incr_0 $
\item Loop to step 2 for a specified number of iterations
\end{enumerate}
\end{minipage}}
\end{center}
\caption{Schematic representation of the ensemble based incremental variational data-assimilation algorithm}
\label{algo-ens4var}
\end{figure}

Relying on the analogy between the recursive Kalman filter update of the 4DVar cost function and a gradient descent procedure, several authors proposed to update the analysis ensemble  through ensemble Kalman filtering equation. This includes approaches based on perturbed observations \cite{Zupansky05} or so-called square-root ensemble Kalman filters \cite{Hunt07}. Local filters have been also considered \cite{Fertig07}. As we saw it previously, these procedures are equivalent only for linear models, and provide the same state only at the end of the assimilation window. Ensemble Kalman filtering techniques provides however a practical mean to obtain the posterior -- also called analysis -- covariance matrix of the estimate given the past observation. In the context of sliding assimilation window, set up for long time integration, we used this ability to construct our initial background ensemble for the next assimilation window. In this work we rely in particular to the perturbed version of the ensemble Kalman filtering formulation of the covariance analysis. Note that square root filters expression could have been as well used here. 



The previous ensemble method relies on a low rank approximation of the background matrix. Such an empirical approximation built from only few samples, compared to the state space dimension, leads in practice to spurious correlations between distant points. For ensemble Kalman techniques, it is customary to remove these long distance correlations through localization procedure. 

\subsection{Localization Issues}
In ensemble Kalman filter, the data assimilation localization are implemented either explicitly, or implicitly.  Implicit localization techniques proceed to observations selection by excluding data points that are located at a distance beyond a distance threshold value. Such localization technique is used in particular for the local filter version \cite{Hunt07}. Explicit localization introduces a Schur - elementwise - product between the background correlation matrix and a local isotropic correlation function:
\begin{equation}
P_b=C \cdot B.
\label{Pb}
\end{equation}
In this work we rely on the latter technique. More specifically, we use a localization mostly based on the work proposed by \cite{Buehner05,  Liu09}. The basic idea is to construct a spatial correlation matrix $C(\|x-y\|/L)$ in which we define a cutoff distance, $L$, and we merely set $C_{xy}=0$ when the distance between $x$ and $y$ exceeds the cutoff value. Gaussian  functions $\exp-\{\|x-y\|^2/2L^2\} $ can be used for that purpose. Polynomial approximations of a Gaussian function with compact support and a hard cutoff are also often employed \cite{Gaspari-Cohn99} as they lead to sparse correlation matrices, which is computationally advantageous. In order to incorporate the localized background error matrix into our system, we need the square root of $P_b$. To do so, a spectral decomposition of this isotropic correlation function 
\begin{equation}
C= E \lambda E^T,
\end{equation}
where $E$ corresponds to Fourier modes, allows us to filter the remaining high frequency components that may lead to erroneous  propagation of spurious high frequency gravity waves generated only by noise. Keeping only the $r$ first leading Fourier modes, we define the approximated correlation square root as:
\begin{equation}
C'=E_{n \times r} \lambda_{r \times r}^{1/2}.
\end{equation}	
The modified perturbation is then provided  by
\begin{equation}
P'_b= (diag(X'_{b,1}) C', \cdots,  diag(X'_{b,N}) C').
\label{Sq_LoBack}
\end{equation}
Here the $diag$ operator sets the vector $X'_{b,k}$ as the diagonal of a matrix.
This localized perturbation matrix will serve us to precondition the assimilation system Eq.\eqref{JincZ-envar-1}, where the forecast background perturbation $\widetilde B\Matsqrt_t$ is,
\begin{equation}
\widetilde B\Matsqrt_t=\tanphi_t  (\inc_0) P^{'}_b.
\end{equation}
Finally $\widetilde B_t\Matsqrt $ can be calculated by
\begin{equation}
\widetilde B_t\Matsqrt = [diag(\tanphi_t  (\inc_0) X'_{b,1}) C', \cdots,   diag(\tanphi_t  (\inc_0) X'_{b,N})C'].
\label{sqB_t_local}
\end{equation}

Let us note this operator can be defined from a general diffusion equation. Compared to convolution approaches, this model is computationally much more efficient and practical to implement on a domain with complex geometry such as in ocean modeling where irregular coastline boundaries are involved. This provides also a method enabling the construction of anisotropic and inhomogeneous correlations. As for its implementation, explicit discretization \cite{Weaver-Courtier01} or implicit discretization have been proposed \cite{Carrier10, Mirouze10}. 

All the elements being now defined, we can then proceed iteratively to the functional minimization. The whole ensemble variational data assimilation technique is summarized in figure \ref{algo-ens4Dvar}.

We already specified the advantages of such a low rank approximation: easy empirical definition from the ensemble fields, no need to use the tangent linear model nor the adjoint operator, control variable's dimensionality reduction. However, the disadvantages are also quite clear: the rank deficiency of $B$ and spurious correlation errors due to the small number of ensemble members. Although long range correlations or erroneous high frequency waves can be alleviated by the localization technique (\eqref{Pb}), the Schur product involved in this procedure may introduce a loss of physical balance between the different variables \cite{Lorenc03, Mitchell02}. Also, the localization technique may lead to underestimate both errors' magnitude and variances in comparison to those derived directly from the ensemble fields \cite{Lorenc03}. An interesting approach to foil these problems, consists to express the background covariance by a balance operator and a diagonal variance matrix learned empirically from the ensemble members \cite{Daget09}. In this approach, the background covariance is formulated as
$B=K D\Matsqrt C D\Matsqrt K\transp$, where $K$ is the linear balance operator upon which the multivariate covariance matrix based is constructed from the univariate diagonal covariance matrix $D$. Matrix $D$ gathers the empirical variance of the background error, it is obtained by transforming the ensemble perturbation through the inverse of $K$; matrix $C$ is a filtering (localization) correlation operator as defined in the previous section. Traditionally, when our assimilation system is exerted on a large scale where the rapid rotation dominates the flow motion, a simple geostrophic relationship can be achieved between the velocity and the pressure gradient \cite{Vidard04}. However in our applications, as the domain of interests is extremely small, the effects of rotation can be ignored, which therefore prohibits the use of the geostrophic relation. In our work, we did not rely on such a technique. However for geophysical application it is clear this decomposition enforces the global physical coherence of the background ensemble.

In the following section we assessed some possible choices of the variational ensemble method scheme and compared their efficiency to a standard incremental 4DVar technique. This experimental evaluation has been carried out for Shallow-water dynamics. 
\begin{figure}[h!]
\begin{center}
\fbox{\begin{minipage}[t]{1\linewidth}
\centering{\em Ensemble  incremental variational data-assimilation algorithm}\\
\begin{enumerate}
\item Set an initial state: $\inc(t_0)=\inc^b(t_0) $ and the ensemble as an arbitrary choice(for the 1st cycle) or as the forecast state and the ensemble forecast derived from the previous assimilation cycle respectively; generate ensemble observation as well
\item Parallelizing compute $\mathbf \inc(t)$ from  $\inc(t_0)$ and each ensemble member $\inc_0^{(p)}$  with the forward integration of the nonlinear dynamics (\ref{Ass-syst-dyn})
\item Derive the background perturbation matrix either by localization technique from (\ref{Sq_LoBack}) (\ref{sqB_t_local}) or (\ref{Sq_EnBack}) (\ref{JincZ-envar-2}) 
 \item Calculate the innovation of the initial state and the ensemble  
 \item Initialized the increment matrix: $\mathbf \incr_0$: [$\incr_0, \incr_0^{(1)}, \cdots ,\incr_0^{(N)}$]
 \item Do an inverse control variable transformation $\mathbf {\delta R_0}= (P'_b)^{-1}\mathbf \incr_0$ if necessary 
 \item Parallelizing optimize $\mathbf {\delta R_0}$ in the inner loop, the cost function and the gradient are calculated based on the modified versions of (\ref{JincZ-envar-1}) and (\ref{GincZ-envar}) respectively
 \item Update control initial space $\mathbf {R_0}$ and calculate $\mathbf \incr_0$  by transforming $\mathbf {\delta R_0}$ to the state space with  (\ref{CVT}) 
\item Update the initial condition: $X(t_0) +:= \incr_0 $, and the initial ensemble $\inc_0^{(p)}$ as well
\item Loop to step 2 for a specified number of iterations
\item Evolve the analysis state $\mathbf{X^a(t_0)}$ to the beginning of the next cycle through the nonlinear dynamics (\ref{Ass-syst-dyn}). The forecast state and forecast ensemble are  used to initialize the next assimilation cycle.
\end{enumerate}
\end{minipage}}
\end{center}
\caption{Schematic representation of the complete ensemble based incremental variational data-assimilation algorithm}
\label{algo-ens4Dvar}
\end{figure} 

\section{Experimental evaluation and comparison}

Before presenting the results,  we first recall briefly the Shallow Water equations and present the experimental setup that we consider in this study.

\subsection{Experimental configuration}
\label{exp-conf}
The experimental configuration that has been chosen consists in a closed tank of 10cm x 25cm x 7cm, filled with water. The tank is tilted along the x direction with a slope of $20\%$ from a horizontal surface. Once the free surface is stable, we lay down the box and start observing the free surface deformation evolution. This experiment was carried out both experimentally and numerically. The figure \ref{aquarium} resumes the experimental procedure:
\begin{figure}[h!]
	\centering
	\includegraphics[scale=0.5]{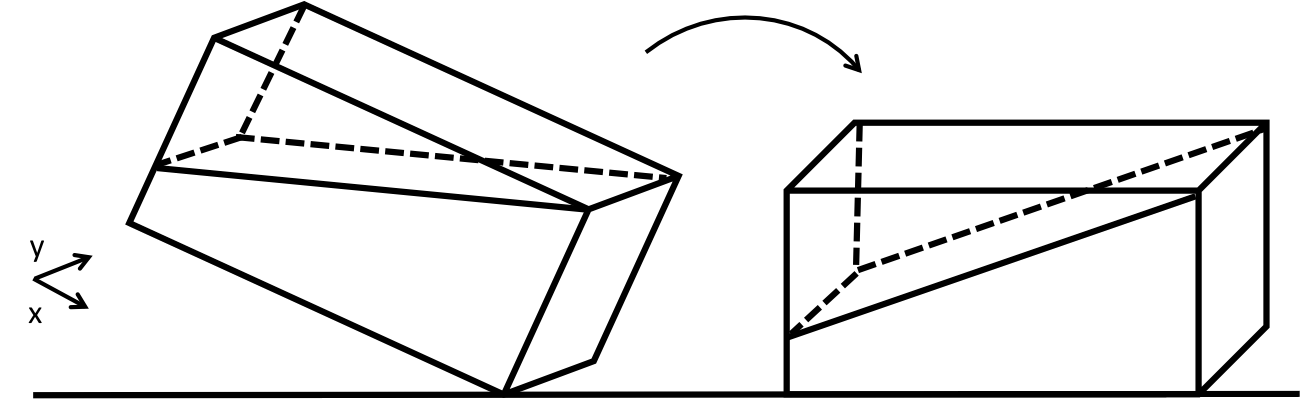}
	\caption{Experimental configuration}
	\label{aquarium}
\end{figure}

Before describing in detail the different study cases we selected, we briefly recall the Shallow-water governing equations on which we rely. 
\subsection{Shallow-water governing equations}
This study focuses on the recovery of a free surface flow for a shallow fluid flow contained in a closed rectangular aquarium. 
This situation can be quite precisely described by the shallow water equations (SWE). These  equations are derived from the following assumption:
\begin{itemize}
	\item The height of the fluid volume  can be neglected compared to its width and length. This implies that we will neglect the vertical shear and vertical acceleration,  and only work with the horizontal velocity component $\bu \defin \bu_H = (u, v)\transp$. 
	\item The pressure is hydrostatically distributed along the vertical, that is, $p = p_0 + \rho_0 g (h-z) $
	\item The fluid is incompressible.
	\item We don't take into account the Coriolis acceleration nor the bottom friction. The bottom of the domain is assumed to be flat and non tilted. 
\end{itemize}
Consequently, the shallow water system is composed of  a height conservation equation and horizontal momentum equations (see for instance \cite{Vreugdenhil94} for a  thorough description of these equations and how they are derived):
\begin{equation}
	\left \lbrace
		\begin{array}{lll}
			\partial_t h + \partial_x(hu) + \partial_y(hv)  & = & 0, \\
			\partial_t (hu) + \partial_x(hu^2) + \partial_y(huv) + \frac{1}{2} g \partial_x h^2 & = & 0, \\
			\partial_t (hv)  + \partial_x(huv) + \partial_y(hv^2) + \frac{1}{2} g \partial_y h^2 & = & 0.
		\end{array}
	\right.
	\label{SWE0}
\end{equation}

Denoting the vector of conserved variables $X = ( h\;hu\; hv)\transp$, the horizontal flux 	$F(X) = (hu  \; hu^2\!+\!\frac{1}{2}g h^2 \; huv )\transp $ and  the vertical flux $G(X) = (hv\; huv\;hv^2\!+\!\frac{1}{2}g h^2)\transp$,  the system \eqref{SWE0} can be compactly written in its conservative form:
\[\partial_t X + \partial_x F(X) +  \partial_y G(X) = 0. \]

The numerical implementation on which we rely is briefly described in appendix \ref{SW-implementation}. The corresponding tangent and adjoint code were constructed with the automatic differentiation tool TAPENADE \cite{Hascoet04}. 


\subsection{Results on numerical synthetic case}
In order to provide some quantitative comparison elements between the different methods, we first carried out experiments on a twin experiment built directly from the Shallow water numerical scheme. The setup of these experiments are detailed in the following section.  

\subsubsection{Twin synthetic experiments}
\label{SyntheticCase}
We set up a twin experiment from two reference trajectories obtained by running the numerical model from two different initial conditions. We then generate noisy observations from an i.i.d. Gaussian noise perturbation of the references' free surfaces height and velocity fields. In the following, we refer to the first synthetic experiment as case A. In this case, we define the reference initial condition as the {\em a priori} initial state of the experimental configuration (i.e. the 20\% tilted initial flat surface)  perturbed with an homogenous Gaussian noise defined by an exponential covariance of decorrelation length (about $5\%$ of lengthy scale) and  variance  ($1.6 mm^2$). This first experiment is thus characterized by a trajectory whose initial state is similar to the {\em a priori} experimental condition up to a Gaussian random field. It is important to underline that within this experiment, we only assume to know this {\em a priori} initial condition. The synthetic true initial condition is completely unknown. As a consequence, we will not be able to provide an ensemble whose mean corresponds to the true synthetic initial condition. 

The second case, referred as case B,  is generated by changing the initial slope of the ideal  free surface. For this case, we fixed the slope in $x$-axis as $21\%$ instead of $20\%$, and we opt for a stronger slope of $10\%$ along the $y$-axis. As in the previous case, only the ideal experimental condition is assumed to be known. Case B deviates strongly more from the ideal case. This case constitutes a good test to assess the ability of the method to correct the initial condition, given a background trajectory that significantly diverges from the true synthetic reference. The observations are generated in the same way as in the first case. 

In both experiments the initial velocity is built for both velocity components from a homogeneous  Gaussian random field with an exponential covariance associated to a standard deviation of $1 mm.s^{-1}$  and a decorrelation length of  $5\%$ of lengthy scale.


The figure \ref{XbXtfig} summarizes the different initial condition configurations considered for the twin experiment:

\begin{figure}[h!]
	\centering
        \includegraphics[scale=0.4]{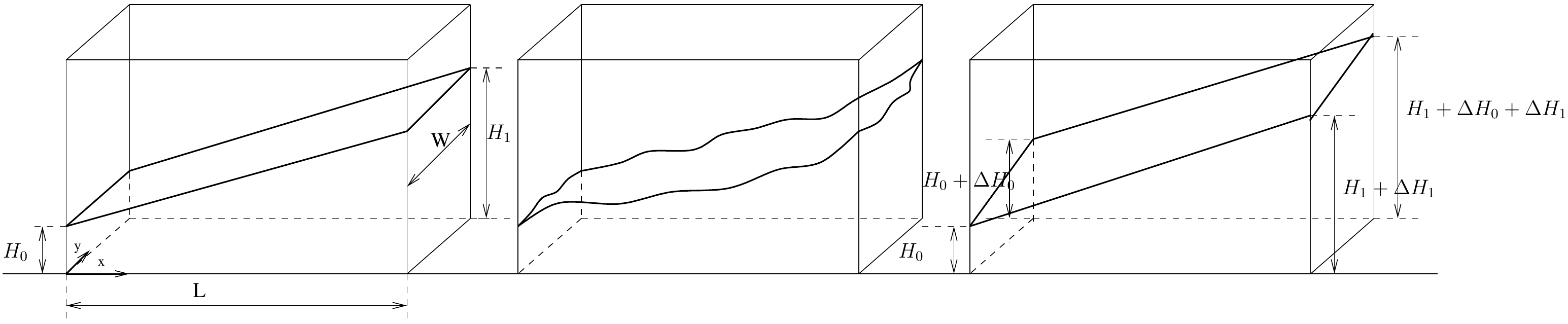}
	\caption{A priori initial experimental configuration (on the left) and the true synthetic initial conditions with a Gaussian noise -- Case A -- (in the middle) and with a $10\%$ slope along the y-axis -- Case B-- (on the right). }
	\label{XbXtfig}
\end{figure}

In addition to these cases, the assimilation schemes were evaluated considering, on one hand, partially observed systems where only the free surface or the velocities are observed, and, on the other hand, fully observed system in which both variables are observed.
In all the experiments, the assimilation window has been fixed to an interval of 0.2 seconds with 5 observations.  The standard deviation of the observation noise have been fixed to $1 mm, 1 mm.s^{-1}, 1 mm.s^{-1}$ for the height and the velocity respectively.


In order to construct the background error covariance matrix, we use different strategies with respect to the incremental 4DVar and the En4DVar techniques. For the incremental 4DVar, we adopt a static diagonal matrix $B=\sigma_b^2 \Id$, where $\sigma_b$ is the standard deviation between the true solution and the {\em a priori} experimental initial state. For the En4DVar, as the background error covariance is derived from the ensemble fields, it is crucial that the initial ensemble represents correctly the background errors. In case A, we generated an initial ensemble perturbing the {\em a priori} initial state with the same i.i.d. Gaussian perturbations of covariance $\sigma_b^2$ as the reference. In case B, apart from the Gaussian perturbation approach, considering that the reference solution deviates strongly from the {\em a priori} configuration, we experimented another strategy which consists in defining initial members from a random drawing of different slopes. Concerning the velocity components, in all the cases, the background free surface velocity is fixed to a null field. Thus, there is a deviation between the reference velocity fields and the known background.
We point out that the initial state (perturbed surface height and null velocity field), is integrated for a few time steps before we start our assimilation process. This provides us the guaranty of balanced velocity perturbations that are conform to the nonlinear dynamic model.

%
%
%
%
%


\subsubsection{Results on case A ( 20\% slope on $x$-axis with additive Gaussian perturbation on the initial surface height)}
In this case, we compared the 4DVar and EnVar on the basis of a partially observed system in which only the velocity measurements are available. As explained above, the initial background matrices were constructed on a similar basis. They both correspond to an i.i.d Gaussian distribution of covariance $\sigma_b^2 \Id$, $\sigma_b$, corresponding to the true deviation between the initial solution of the reference and the given {\em a priori} background. The background error is however biased as its mean is unknown. 
%
%
%

\begin{figure}[h!]
\centering
\includegraphics[width=1\textwidth]{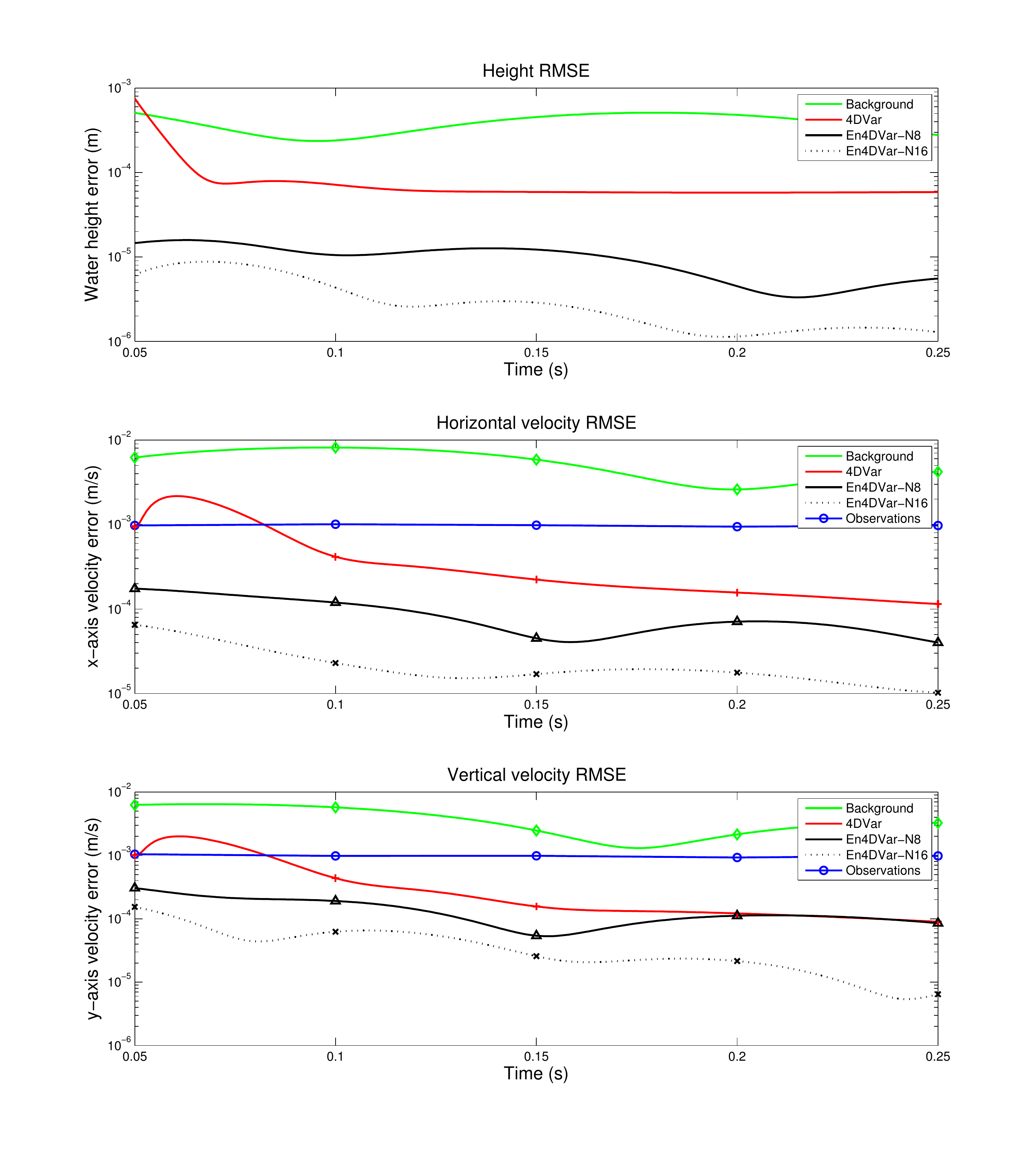}
\caption{RMSE comparison between an incremental 4DVar and the En-4D-Var assimilation on a coarse resolution grid in semi-log plot axis -- system observed through  noisy velocity fields.}
\label{obsVel_caseA_low_loglog}
\end{figure}

The tests were carried out for two different resolutions to assess  the importance of the localization technique used in En4DVar. The assimilation schemes were both run on a coarse $11\times26$ resolution grid and no localization was used in the background covariance definition of  En4DVar. Figure \ref{obsVel_caseA_low_loglog} pictures for both assimilation techniques the root mean square error (RMSE) curves of the reconstructed free surface curves and its associated velocity component. These curves  show that even with a low number of ensemble members $N=8$, En4DVar provides better results than the standard incremental 4DVar technique. The ensemble  approximation of the time evolving background covariance introduced {\em de facto} by the ensemble technique seems to be quite efficient compared to the fixed background covariance of the standard 4DVar. For 16 members, the ensemble technique significantly outperforms the standard 4DVar technique. 

\begin{figure}[h!]
\centering
\includegraphics[width=1.\textwidth]{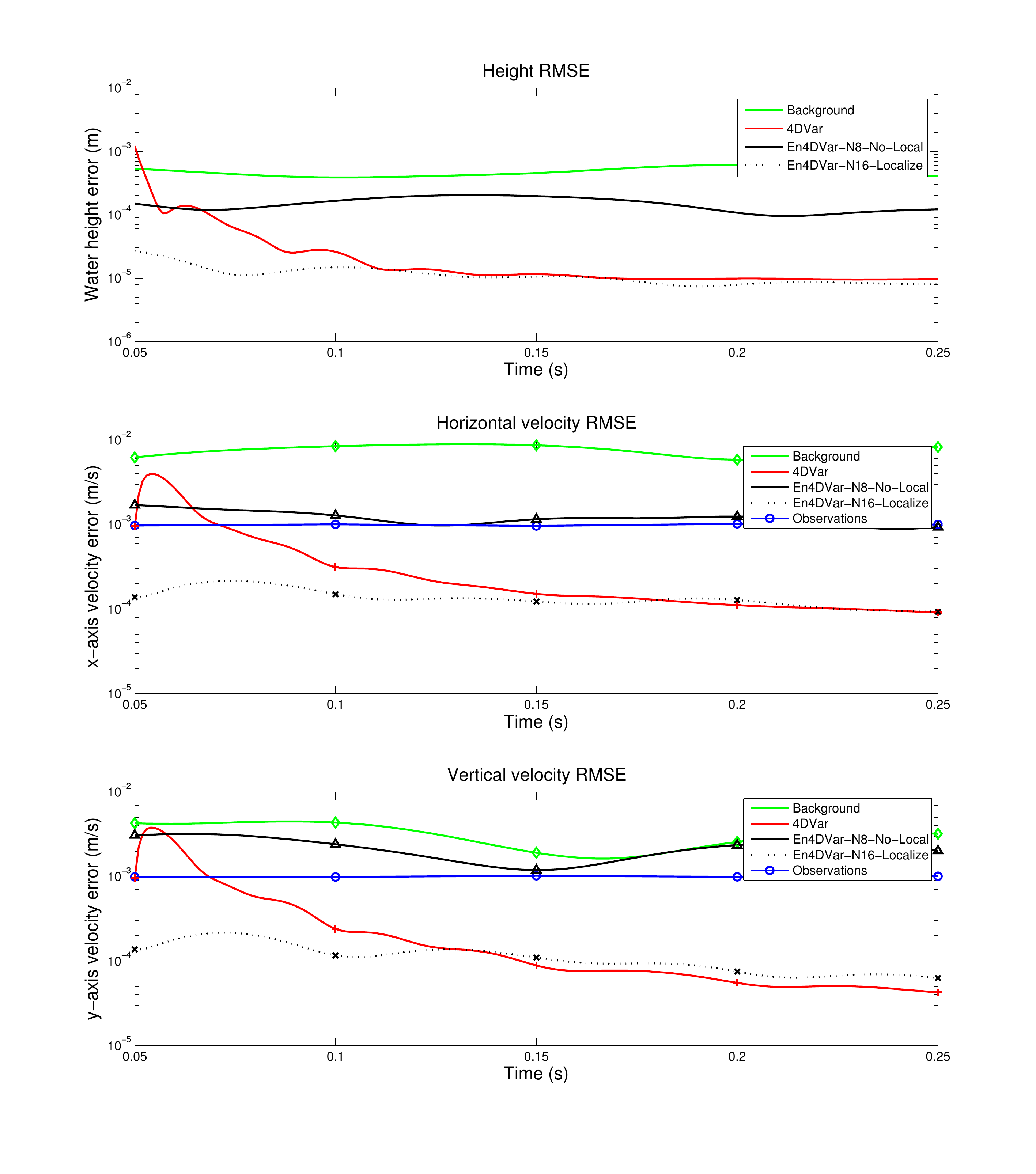}
\caption{RMSE comparison between an incremental 4DVar and En-4D-Var assimilation on a fine 
resolution grid in semi-log plot axis -- system observed through noisy velocity fields.}
\label{obsVel_caseA_loglog}
\end{figure}

For the finer resolution of the model,  the results shown in figure \ref{obsVel_caseA_loglog} reveals  the  necessity of localizing the empirical error covariance. Without the use of a localization filter, the standard 4Dvar with a static background provides clearly better results. With  localization, both assimilation techniques yield similar quality results in terms of the surface height or velocity components RMSE. High resolution amplifies the ensemble spread and brings out spurious long range correlation factor between far apart grid points. As already stated in several studies \cite{Houtekamer01, Fairbairn13}, for a small ensemble size, localization improves clearly the ensemble performances.  This example shows that this necessity is also related to the model resolution grid. At a coarse scale, an ensemble of only few samples allows a good representation of the background error statistics without any localization.  

\subsubsection{Results on Case B: (21\% slope on $x$-axis  and  $10\%$ slope on $y$-axis)}
For the case B, the assimilation techniques were evaluated with partially observed systems where only the free surface height or the velocity components are observable, and also a fully observed case in which measurements for the whole system are available. The results  corresponding  to each case are reported in the subsequent paragraphs. 

Let us recall that in this case, the initial condition of the reference is far apart from the known {\em a priori} configuration fixed as the background -- without slopes in the $y$ coordinate. As explained previously we tested two different initial background covariance setups. The first one is a Gaussian random field perturbation as in the previous case. This case is indicated with a "Gauss" suffix in the result curves. The second background covariance configuration is built from a random sampling of the initial surface with different slopes.  The mean is taken, as always, to be the (incorrect) {\em a priori} background.  This case is indicated by the "Para" suffix. The experiments were conducted for different sizes of the ensemble: 8 members and 16 members are respectively represented by a solid line and a dash line; when necessary, we also displayed 32 members by a dot line among the RMSE curves.

\begin{figure}[h!]
	\centering
	\includegraphics[width=1\textwidth]{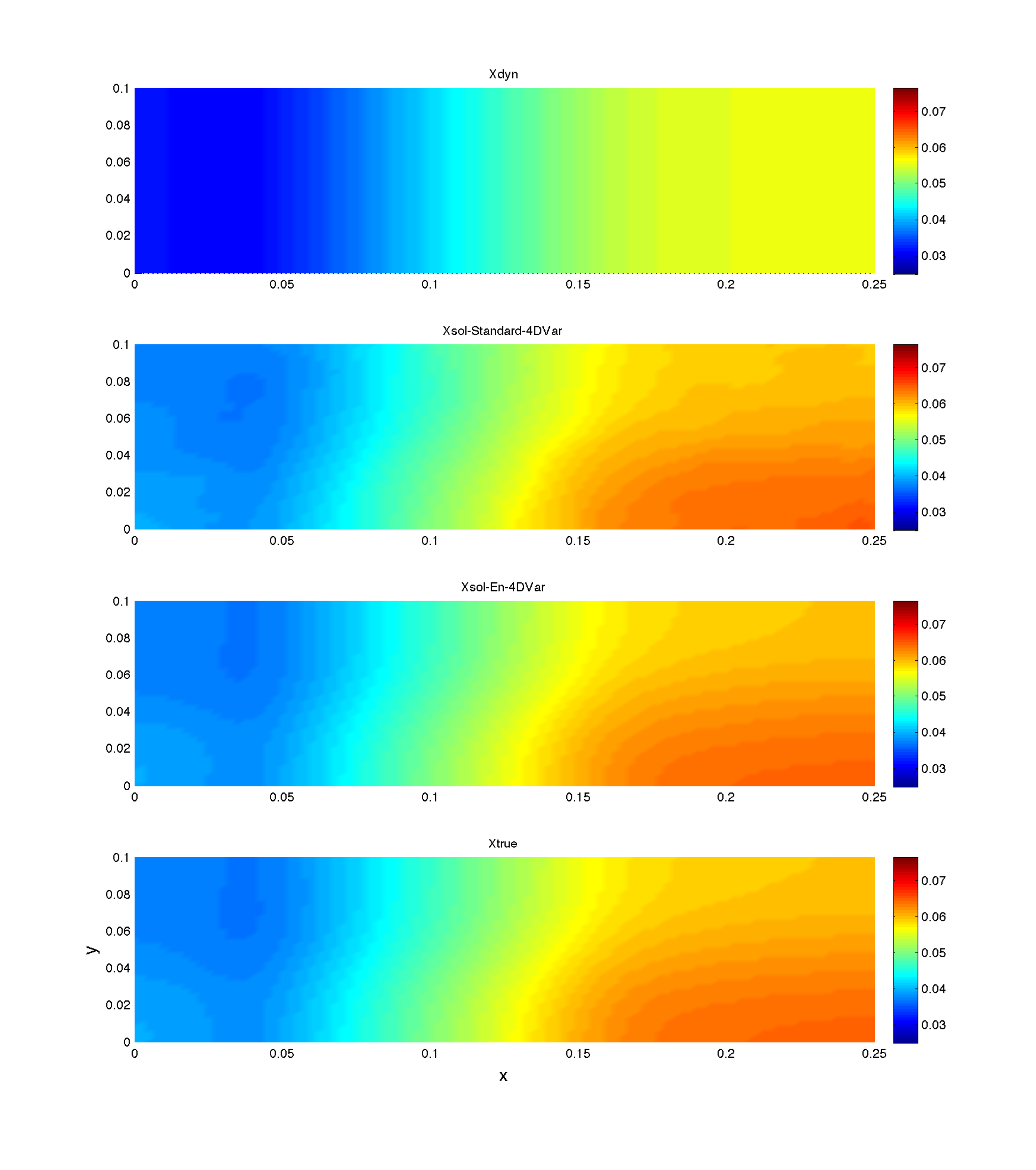}
	\caption{Height field comparison at the middle of the assimilation window, from top to below background, incremental 4DVar, En-4D-Var(N=16), true solution}
	\label{obsH_b10_hei}
\end{figure}

\paragraph{Height observation}
\begin{figure}[h!]
	\centering
	\includegraphics[width=1\textwidth]{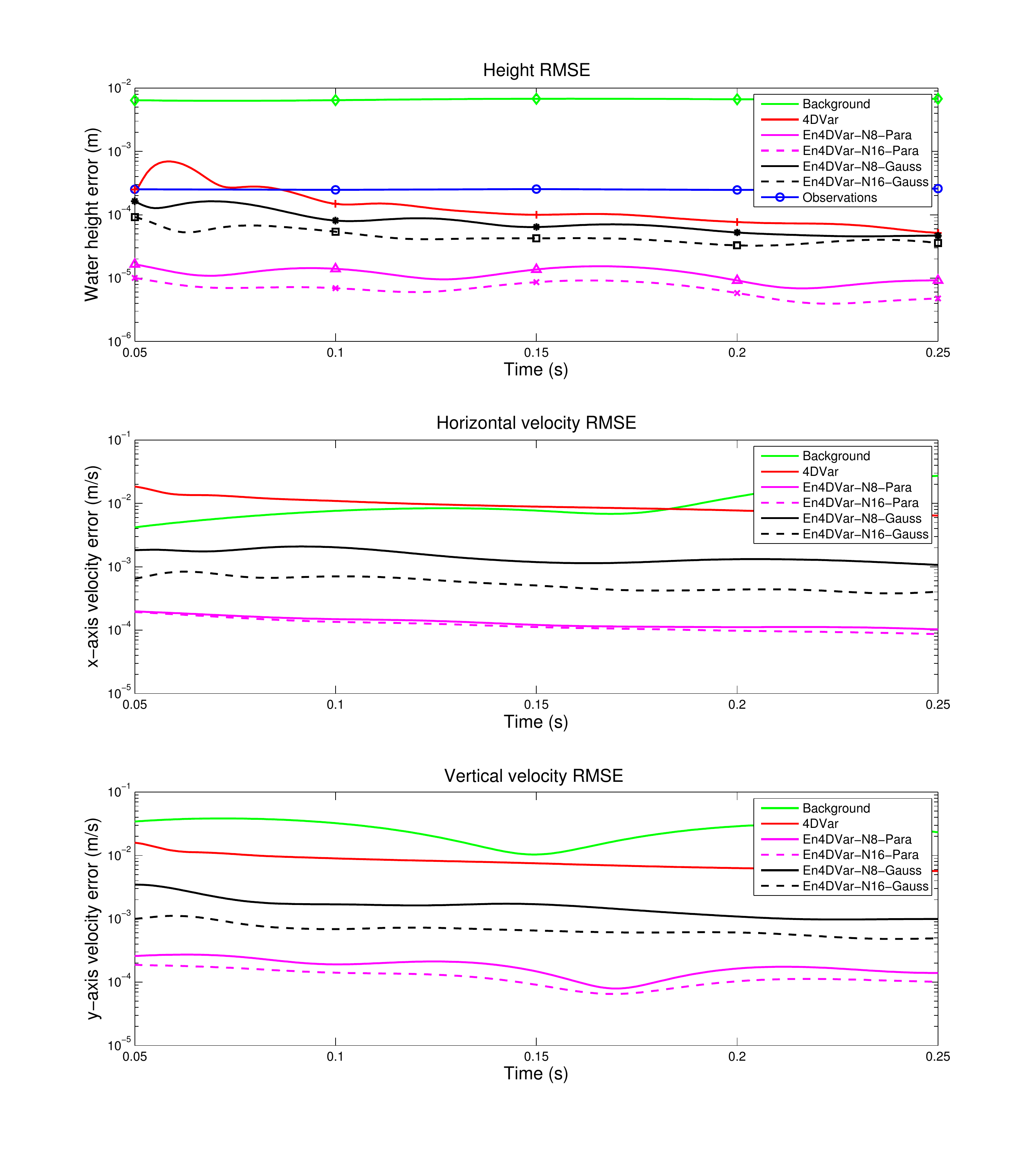}
	\caption{RMS comparison between an incremental 4DVar and En4DVar assimilation semi-log plot -- system observed through noisy free surface height.}
	\label{obsH_b10_loglog}
\end{figure}
The RMSE curves corresponding to this case are gathered in figure \ref{obsH_b10_loglog}. 
We observe that the En4DVar assimilation technique globally leads to better results than the standard 4DVar assimilation technique. The initial free surface is strongly corrected in both methods, however the unobserved velocity components are  well corrected only for the ensemble technique. Note that the mechanisms that come into play for the reconstruction of unobserved components are different in both cases. For standard 4DVar, the adjoint operator of the dynamic model (tangent and adjoint operators for incremental case) propagates a balanced increment from the height to the velocities; for EN4DVar, the correlation terms in the background error covariance, derived from the initial ensemble, play an important role in reconstructing the analysis velocity fields. For the background covariance constructed from a Gaussian field perturbation, the increase of the ensemble size improves the results whereas, in the second case, the results are already of a very good quality for an ensemble of 8 members. A localization procedure was generally needed in the former case but not in the latter. We recall that the localization impact decreases as the ensemble member increases, indeed, as shown by the results associated with ensemble member, more than 32 members (not shown here); also the optimum cutoff distance minimizing the average RMSE exists and its value increases as the ensemble member increases. Both results are consistent with \cite{Houtekamer01, Fairbairn13}. Note that the optimum cutoff distance is always used to illustrate the results. This result highlights the gain of performance that can be achieved by fixing an adequate physical perturbation in the ensemble assimilation. For a  physically adapted perturbation, the performance is improved for all the components, whether observed or not; the ensemble size can also be significantly reduced and no localization appears to be necessary. A set of surface height  recovered by the different methods is displayed in figure \ref{obsH_b10_hei}. 

\begin{figure}[h!]
	\centering
	\includegraphics[width=1\textwidth]{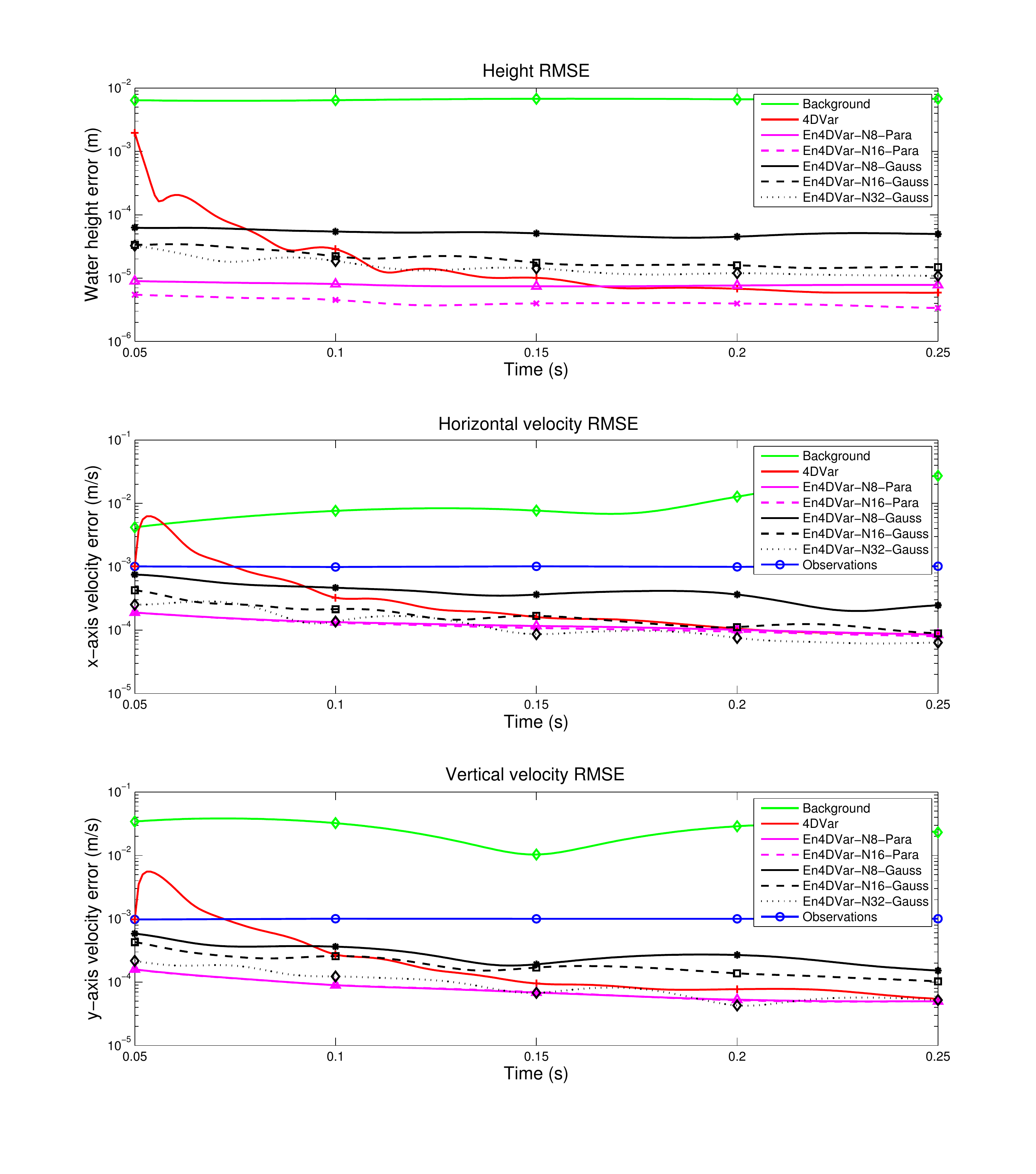}
	\caption{RMS comparison between incremental  4DVar and En4DVar assimilation semi-log plot -- system observed through noisy velocity fields.}
	\label{obsV_b10_loglog}
\end{figure}

\begin{figure}[h!]
	\centering
	\includegraphics[width=1\textwidth]{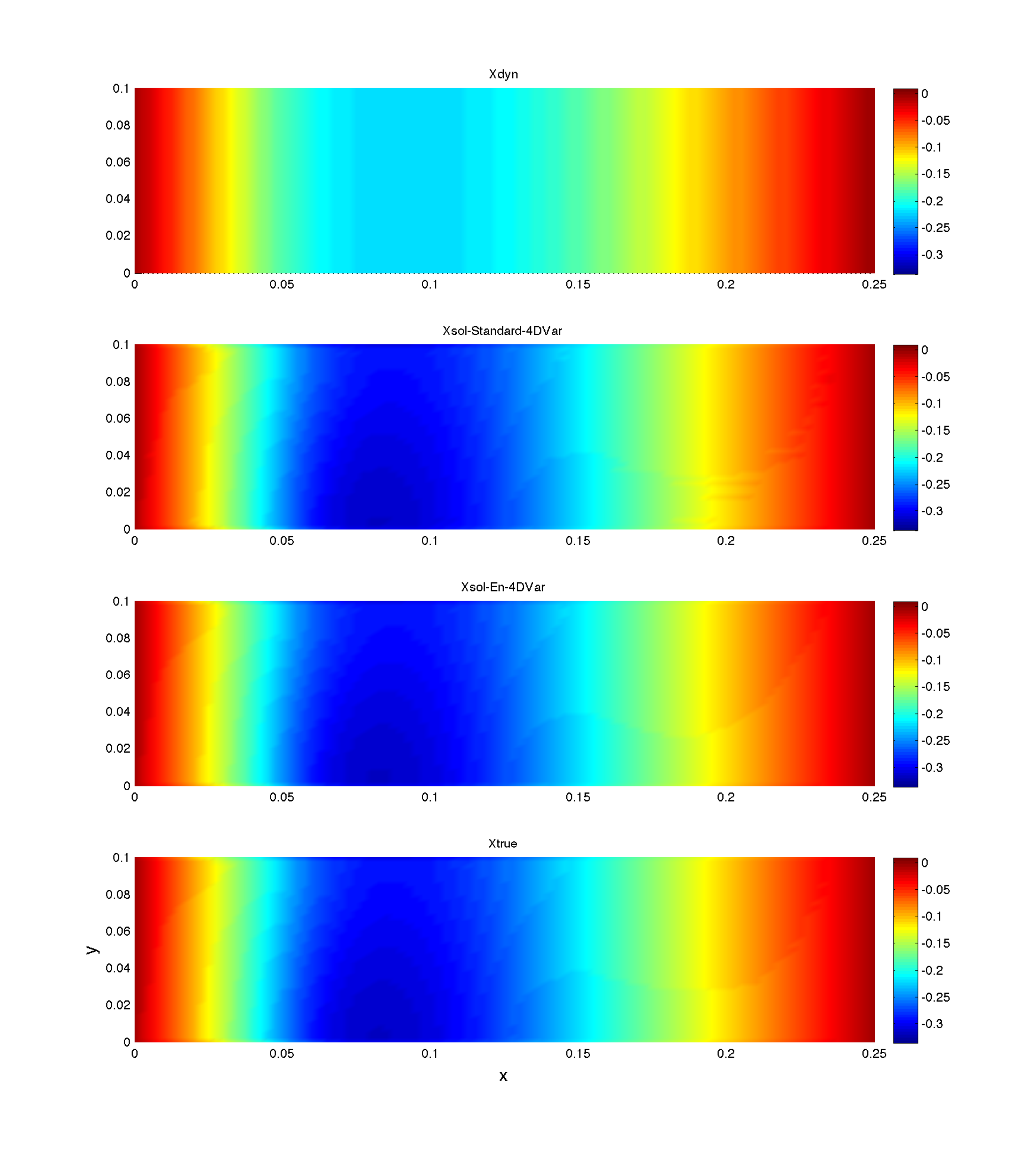}
	\caption{Horizontal velocity field comparison at the middle of the assimilation window, from top to below background, incremental 4DVar, En4DVar(N=32), true solution}
	\label{obsV_b10_Uvel}
\end{figure}

\begin{figure}[h!]
	\centering
	\includegraphics[width=1\textwidth]{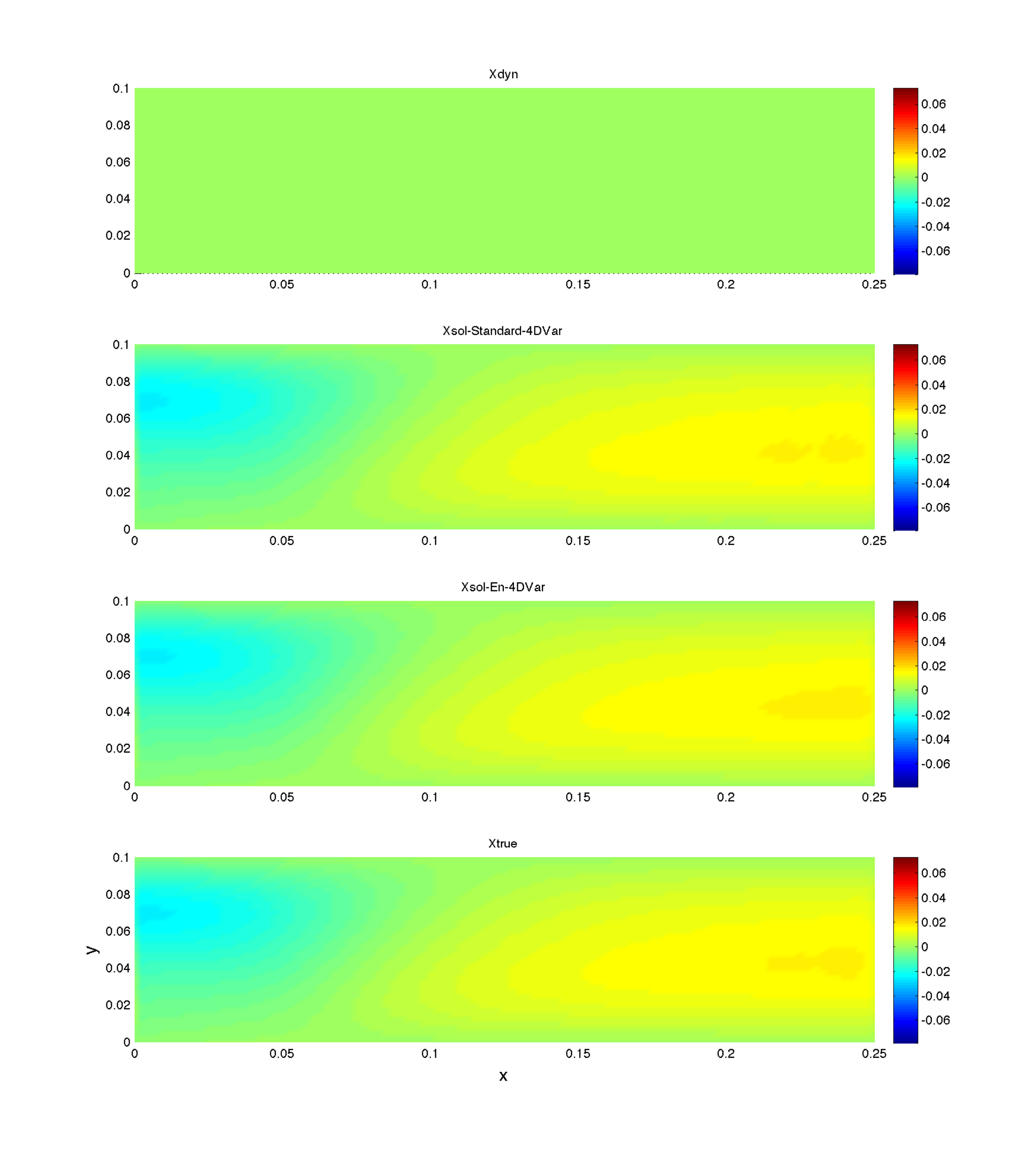}
	\caption{Vertical velocity field comparison at the middle of the assimilation window, from top to below background, incremental 4DVar, En4DVar(N=32), true solution}
	\label{obsV_b10_Vvel}
\end{figure}

\paragraph{Velocity observations} 
The results corresponding to this case are shown in figure \ref{obsV_b10_loglog}. In this case, we retrieved almost the same behavior as in the previous case. When associated to an unphysical background covariance built from a Gaussian perturbation, the En4DVar techniques require to rise the size of the ensemble in order to obtain results of comparable quality to the standard 4DVar. A localization filter is needed in that case, and a too low number of members leads to weaker results. Compared to ensemble techniques, the incremental 4DVar technique shows some difficulty to correct efficiently the unobserved surface height component. The En4DVar, with an initial ensemble built from random slope parameters \DEL{drawing}, leads to better results with a moderate number of members ($N=16$ in this case). A lower number ($N=8$) yields good results for the observed variables but not for the unobserved surface height. An increase of the ensemble size ($N=32$) slightly improves the quality of the results. Some velocity fields recovered by the different techniques are shown in figures \ref{obsV_b10_Uvel} and \ref{obsV_b10_Vvel}.

In terms of the computation time, based on our platform (2 Quad-Core Intel Xeon, 8 Processors,16GB memory available), the En4DVar method takes 4 times more computation time when doubling the ensemble members using moderate localization. While the 16 members case still consumes a similar time span compared to the standard 4DVar, the 32 members case requires a tremendous computation time. However, the computation time can be largely reduced by the parallelization of  the ensemble. Most of the computation consumption of the the standard 4DVar is dedicated to the adjoint model, whilst En4DVar demands more RAM space due to its time varying background error covariance matrix representation. Once the latter background error covariance is efficiently stored, the calculation of the optimization routine is quite straightforward. 



\paragraph{Height and velocity observations}

Finally, we compared the assimilation techniques assuming a fully observable system where measurements of both height and velocity fields are provided. Similarly to the previous case, the ensemble technique with the background covariance obtained from a Gaussian perturbation requires the augmentation of the ensemble size to obtain comparable results to those obtained with the standard 4DVar.  A smaller ensemble leads to worse results and the use of a localization filter is necessary. The results obtained for the physical background covariance are slightly better than the results obtained for 4DVar. This good behavior can be observed even for a small number of members. Nevertheless, the advantage of ensemble methods are less enhanced than in the previous cases. This is probably due to the fact that under this circumstance, each variable component is mainly corrected by the corresponding observations rather than indirectly from other observed components. However, in real world applications, assimilation problem are rarely fully observable, hence the capacity of ensemble methods to correct efficiently the unobserved system's components appear to be very attractive. 





\begin{figure}[h!]
	\centering
	\includegraphics[width=1\textwidth]{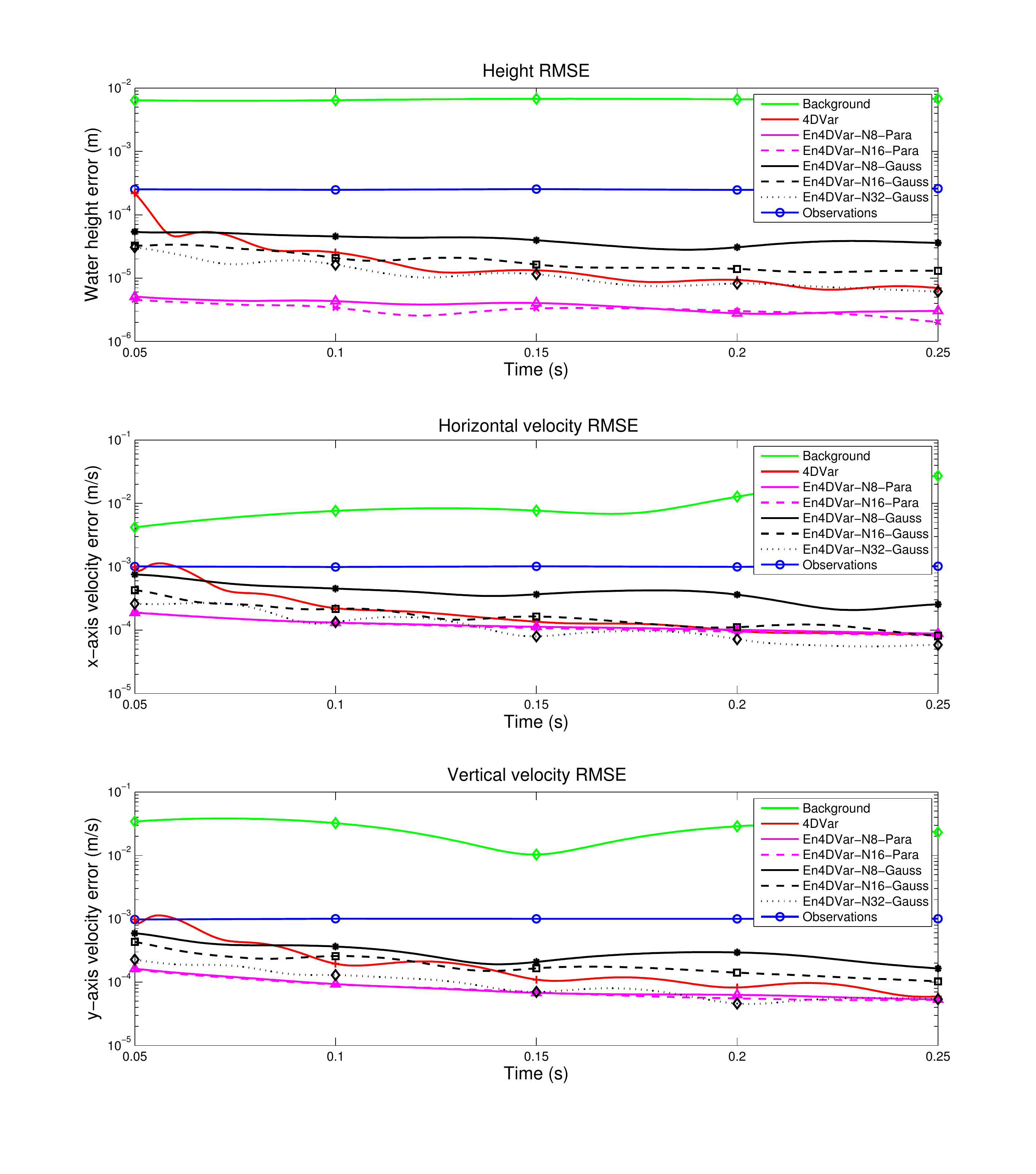}
	\caption{RMS comparison between incremental 4DVar and En-4D-Var assimilation and its semi-log plot -- fully observed system through noisy velocity fields and noisy free surface height.}
	\label{obsAll_b10_loglog}
\end{figure}

%

\subsection{Result on an experimental case}

\begin{figure}[h!]
        \centering
        \includegraphics[scale=0.5]{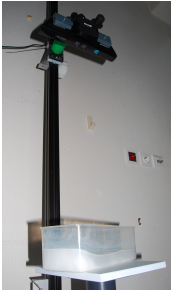}
        \caption{Experimental set with the Kinect sensor.}
        \label{kinect}
\end{figure}

We carried out a last evaluation on a real world experiment that nearly corresponds to the conditions described in section \ref{exp-conf}. Figure \ref{kinect} shows a picture of the experimental setup.
In this experiment context, the measurement corresponds to the surface height provided by the means of the depth sensor (Kinect sensor figure \ref{kinect}). These observations are characterized by a high level of noise and exhibit large regions of missing data all over the boarders because of light reflections on the tank's wall.   In order to cope with these incomplete observations, within the unobserved region, the error is set to a maximal value with a factor that depends on the distance from the closest observed point. The longer the distance is, the larger the error will be. For the other observed points, the instrument error is set to $\sigma_o=0.5 mm$.

\begin{figure}[h!]
\centering
 \subfloat[][]{\includegraphics[scale=0.3]{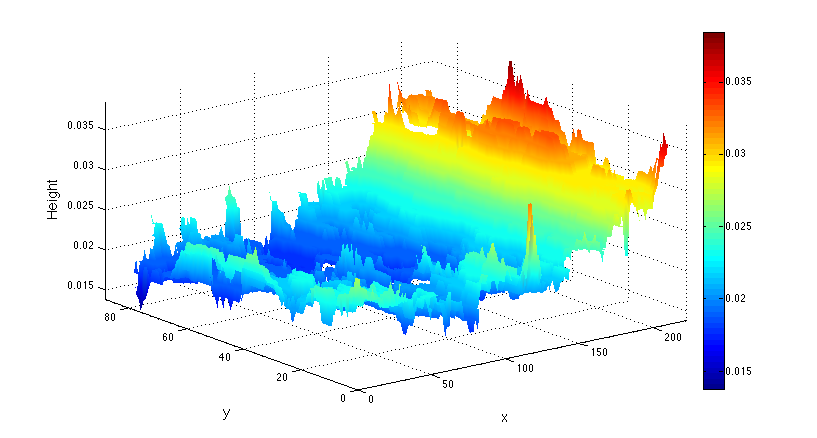}\label{obsexp}}
 
 \subfloat[][]{\includegraphics[scale=0.3]{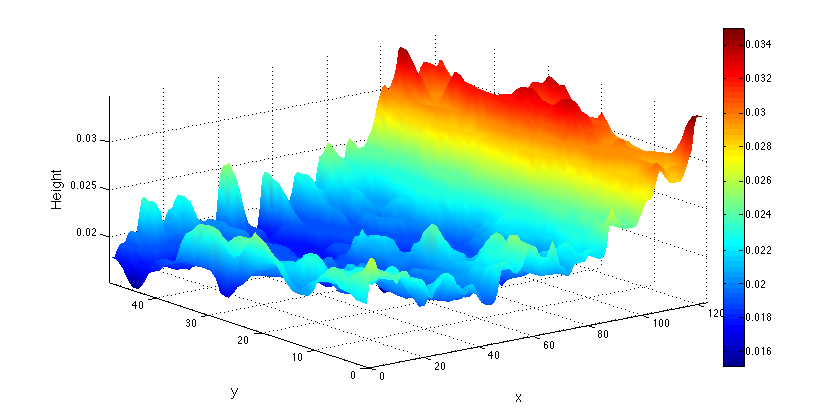}\label{xbexp}} 
 
 \subfloat[][]{\includegraphics[scale=0.3]{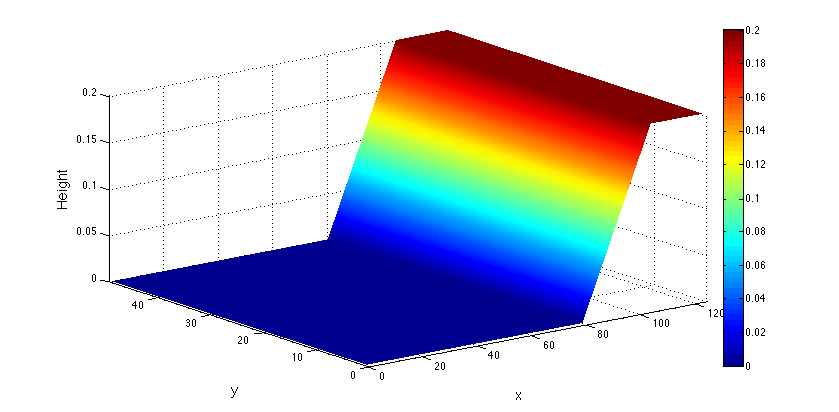}\label{xbVexp}} 
\caption{The height field observed from the kinect camera \protect\subref{obsexp}, the corresponding height background state \protect\subref{xbexp} and primary velocity magnitude \protect\subref{xbVexp} at $t=0$. } 
\end{figure}

For this experimental case, the background is completely unknown and was set to the observed height field at the initial time.
More precisely, the background initial height corresponds to a filtered observation with interpolated values on the missing data regions. The initial time corresponds to the instant at which the tank is lain on the horizontal surface. The figures \ref{obsexp} and \ref{xbexp} show the first observation and the corresponding interpolated and filtered initial state respectively. Concerning the initial velocity field, we roughly set it in order to correspond to the profile of the free surface height gradient, as showing in figure \ref{xbVexp}.

Similarly to the synthetic case, the assimilation starts at the second image in order to construct balanced ensemble through the integration between the two first images of a set of members defined by an i.i.d. Gaussian perturbation with standard deviation $\sigma_b=1.8mm$. In this case, the true solution is  completely unknown. The assimilation scheme was adapted to sliding assimilation windows. The goal of setting multiple windows is to avoid long range temporal correlations. We adopted for 5 windows over 9 observation times, each window consists in 5 observations and each window starts at the 1st, 2nd, ... , 5th observation respectively.

\begin{figure}[h!]
        \centering
        \includegraphics[width=0.8\textwidth]{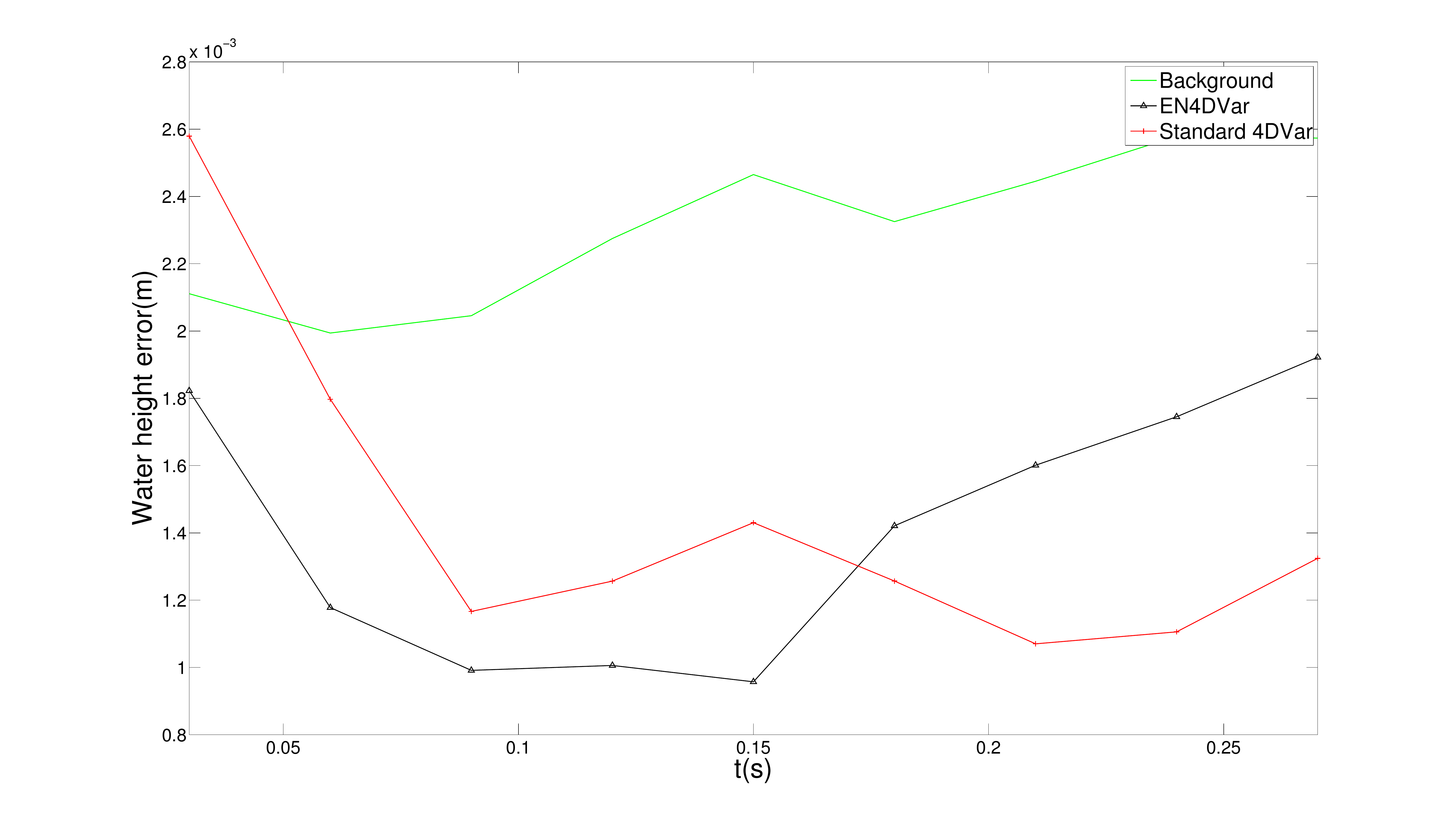}
        \caption{RMSE comparison between classic 4DVar and En-4D-Var assimilation plot.}
        \label{obsH_exp_loglog}
\end{figure}

The results obtained by both assimilation techniques are displayed in figure  \ref{obsH_exp_loglog} in terms of the RMSE between the observed surface height and the estimated one. The RMSE between the background solution and the observation is also plotted on this figure for comparison purpose. We can observe that the 4DVar and the En4DVar lead both to a lower RMSE than the background solution. Note that the RMSE of the 4DVar at the initial time is however higher than the background.  

\begin{figure}[h!]
        \centering
        \includegraphics[width=1\textwidth]{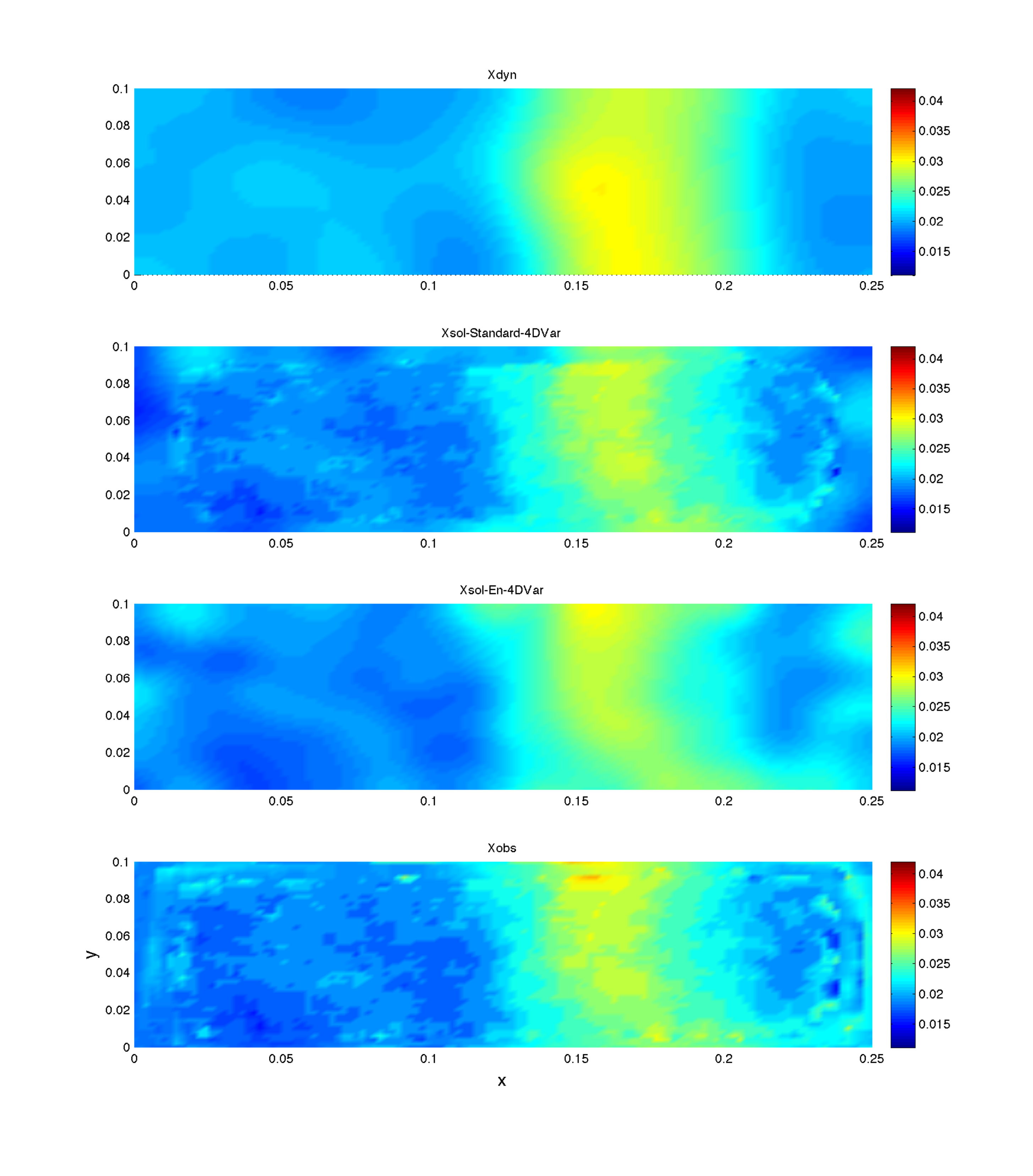}
        \caption{Height field comparison at the middle of the assimilation windows, from top to below background, standard 4DVar, En-4D-Var(N=16), observation}
        \label{obsH_exp_hei}
\end{figure}

We also compared the free surface in figure \ref{obsH_exp_hei}. According to these free surfaces, we can see that the 4DVar solution has some difficulties to handle the discontinuities at the boundaries of the regions in which the data have been extrapolated. Discontinuities in the 4DVar solution between the observed regions and the very noisy region appear clearly. The En4DVar provides a much more satisfying results on the boarders. They are smoother and correspond clearly to a better compromise between the observation and the model.
 
\section{Conclusions and perspectives}
In this paper, an ensemble variational data assimilation technique has been proposed and evaluated within the framework of a 2D shallow water model. The results were compared with a standard 4DVar technique with both synthetic and real observations deduced from images. Such an ensemble variational technique corresponds to a relatively new method. Its advantages lay in its implementation which does not involve the adjoint of the dynamics tangent linear operator and the handling of a flow-dependent background error matrix that can  be consistently adjusted to  the background error. 

However, due to the limitation of the sampling numbers, the rank of the background error covariance has to be largely reduced, which suggests that the analysis is obtained rather on a subspace of much lower dimension than the original model space. This can be the source of pseudo correlations and even worse, it can lead to an unsolvable optimization problem. This limitation can be alleviated by localizing the background error matrix which increases the rank of the background error covariance at the risk of higher computational needs. With respect to our small domain problem, localization is specially effective when dealing with a small ensemble. An adequate localization has the same effect as doubling the ensemble which demands more computational power. 

As for the results that come out from the evaluation we performed, in a nutshell, when observing all the variables involved in the physical model, the ensemble method can achieve slightly better results than the incremental 4DVar technique at the expense of a higher computational cost. When components of the state variables are unobserved, the ensemble technique  outperforms the standard 4DVar for a smaller computational cost. This is quite interesting as in real world applications, it is usually impossible to access to all the variables at hand. We showed in particular that the empirical multivariate correlations, if properly posed, can greatly improve the quality of the non-observed analysis.

At last, the premise of using ensemble variational assimilation to achieve comparable performances with  incremental standard method is that the initial ensemble of the assimilation process has to fit  the physics of the observed phenomena. 

Future work will apply the method to directly assimilate image data by a surface quasi-geostrophic (SQG) model when assimilating the surface temperature and velocities which should have more quasi-operational implications. Also, we try to shift to a stochastic shallow-water model \cite{Memin13}, as in the framework of image assimilation, the resolution of observation space is normally quite high, the use of a stochastic model could be computationally more advantageous by carrying out the dynamics on a coarser grid instead of keeping the subgrid stress tensor.

\newpage
\appendix
\section{Variational data assimilation}\label{assim}
In this appendix we present briefly the setting of the adjoint optimization strategy of the objectif function (\ref{Jnoinc}) subject to dynamical system (\ref{Ass-syst-dyn})  describing the evolution of the state variables with the initial condition (\ref{Ass-syst-init}). Formally, we assume that the state variable 
{\small $\X(t)\in \hil$} and  the observation {\small $\obse(t) \in \STobs$}  are square integrable functions in Hilbert spaces
identified to their dual. The norms corresponds to the Mahalanobis distance  defined from the inner products {\small $<R^{-1}.,.>_\STobs$}, {\small $<B^{-1}.,.>_\hil$}  of the measurements and  the state variable  spaces respectively. They involve covariance tensors  $R$, and $B$  related to the measurement error and the error on the initial condition.  

In order to compute the gradient of this functional, we assume that
{\small $\X(u(t),\noise;t)$} depends continuously on $\noise$ and is
differentiable with respect to the control variables $\noise$,
 on the whole time range. We also assume that the dynamical system admits a solution on the assimilation window time range.
 \subsection{Differentiation}
First of all, given {\small $d\inc = (\partial \inc/\partial\noise)  \delta
\noise$}, the differentiation of equations
(\ref{Ass-syst-dyn}--\ref{Ass-syst-init}) in the direction
$\delta\noise$ is given by:
 \begin{eqnarray}
\label{eq:diff_dyn}
    \partial_t d\inc+ \partial_{\inc} \mdl (\inc) d\inc&=& 0,  \\
\label{eq:diff_ini}
    d\inc(\x,t_0) &=& \delta\noise(\x).
 \end{eqnarray}
The differentiation of the cost
function (\ref{Jnoinc}) in the direction $\delta\noise$ is then:
\begin{equation}
\Big<\frac{\partial {\cal J}}{\partial \noise},\delta\noise\Big>_\hil =\Big <(\inc(\x,t_0)-\inc_0(\x)),\delta\noise\Big>_{\hil}-\\ \int_{t_0}^{t_f}\Big < \obse(t) -
\mobs(\inc(t)),\tanH(\frac{\partial\inc}{\partial \noise}
\delta\noise)\Big>_{\STnoiseo}dt.
\end{equation}
\noindent
Introducing the adjoint  of the linear tangent operator $\adjH$, defined as:
\begin{equation}
\forall (x,y) \in (\hil,\STnoiseo), <\left(\partial_{\inc}\mobs\right)x,y>_\STnoiseo = <x,\adjH y>_\hil,
\end{equation}
this relation can be reformulated as:
\begin{multline}
\Big<\frac{\partial {\cal J}}{\partial \noise},\delta\noise\Big>_\hil
= \Big <B^{-1}(\inc(\x,t_0)-\inc_0(\x)),\delta\noise\Big>_\hil- \\ \int_{t_0}^{t_f}\Big< \adjH\!\!R^{-1}(\obse(t)-\mobs(\inc(t)),
\frac{\partial\inc}{\partial \noise}\delta\noise\Big>_\hil dt.
\label{eq:diff_adj}
\end{multline}
\noindent This expression constitutes the functional
gradients in the directions $\delta\noise$. A direct numerical evaluation of
the latter expression is in practice too expensive as it requires successive integration of the dynamics along the different initial condition components.
\subsection{Adjoint model}
An elegant alternative consists in relying on an
adjoint formulation \cite{Ledimet86,Lions71}.
To that end, the integration over the range $[t_0,t_f]$ of the inner product between  an adjoint
variable $\adj \in \STinc$ and relation (\ref{eq:diff_dyn})  is performed:
\begin{displaymath}
\int_{t_0}^{t_f}\hspace{-0.15cm}\big< \frac{\partial d\inc}{\partial
  t}(t),\adj(t)\big >_{\hil}dt +
\int_{t_0}^{t_f}\hspace{-0.15cm}\big<\tanM d\inc(t),\adj(t)\big>_{\hil}dt  = 0.
\end{displaymath}
\noindent
 An integration by parts of the first term yields:
\begin{equation}
-\int_{t_0}^{t_f}\big< -\frac{\partial \adj}{\partial
  t}(t))+\adjM \adj(t),d\inc(t)\big >_{\hil}dt = \big<\adj(t_f),d\inc(t_f)\big >_{\hil} -\big<\adj(t_0),d\inc(t_0)\big >_{\hil} ,
\label{eq:int_part}
\end{equation}
where the adjoint of the tangent linear operator
 $\adjM: \hil \rightarrow \hil$ 
has been introduced. At this point no particular assumptions nor constraints were imposed on the adjoint variable. 
However, we are free to specify the set of adjoint variables of interest by setting a particular evolution equation or a given
boundary conditions simplifying  the computation of the functional gradient. 
As we will see, imposing the
adjoint variable $\adj$ as the solution of the system
\begin{equation}
\left\{
\begin{array}{l}
-\partial_t \adj(t)+\adjM \adj(t) = \adjH R^{-1}(\obse-\mobs(\inc(t)))\\
\;\adj(t_f) = 0,
\end{array}
\right.
\label{eq:adj}
\end{equation}
\noindent
will provide us a simple and accessible solution for the functional gradient. 

As a matter of fact, injecting  this relation into equation (\ref{eq:int_part}) with {\small $d \inc(t_0)= \delta\noise$}
and {\small $d\inc =   (\partial \inc/\partial\noise)  \delta \noise$}  allows us to identify the right hand second terms of the functional gradients (\ref{eq:diff_adj}) and we get: 
\begin{displaymath}
\Big<\frac{\partial {\cal J}}{\partial
   \noise},\delta\noise\Big>_\hil\;= -\Big<\adj(t_0),\delta\noise\Big>_\hil+ \Big<B^{-1}(\inc(t_0)-\inc_0),\delta\noise\Big>_\hil.
\end{displaymath}
\noindent
From these relations, one can now readily identify the expression
of the cost function derivative with respect to the control variable:\vspace{-0.05cm}
\begin{equation}
\frac{\partial {\cal J}}{\partial    \noise} =  -\adj(t_0) + B^{-1}(\inc(t_0)-\inc_0).
\label{eq:gradj}
\end{equation}

\noindent The partial derivatives of ${\cal J}$ are now simple to
compute when the adjoint variable $\adj$ is available. The knowledge of the functional gradient 
allows us to define updating rules for the control variables from iterative optimization procedures. A quasi-Newton  
minimization process consists for instance of:
\begin{equation}
\inc_{n+1}(t_0)= \inc_n(t_0) - \alpha_n \tilde{H}^{-1}_{\inc_n(t_0)}( B^{-1}(\inc_n(t_0)-\inc_0) -\adj(t_0) ),  
\label{eq:gradj2}
\end{equation}
where $\tilde{H}^{-1}_{x_n}$ denotes an approximation of the Hessian inverse computed from the functional gradient with respect to  variable $x_n$ and  the constant $\alpha_n$ optimally chosen.

\subsection{Shallow water numerical scheme}
\label{SW-implementation}
Hereafter, we briefly describe the numerical implementation of the Shallow Water Equations (SWE) on which we rely in this study. Without loss of generality, we only consider the following 1D version of SWE:
\begin{equation}
	\left \lbrace
		\begin{array}{lll}
			\partial_t h + \partial_x(hu)   & = & 0, \\
			\partial_t (hu) + \partial_x(hu^2 + \frac{1}{2} g  h^2) & = & 0. \\
		\end{array}
	\right.
	\label{SWE1D}
\end{equation}
This system can be written in a conservative form 
\begin{equation}
	\partial X_t(t,x) + \partial_x F(X(t,x))  = 0,
	\label{conservativeNL}
\end{equation}
where $X =(h \; hu)\transp $ and 	$F(X) = (hu\;hu^2 + \frac{1}{2}g h^2)\transp$ denote respectively the conserved variables and their associated  fluxes. 

If the $\partial_x F$ operator is linear and diagonalisable, the solution is straightforward. We first perform a diagonalisation $\partial_x F = R^{-1}\Lambda R$, with $\Lambda$ the diagonal matrix gathering the  eigenvalues $e_1 < \cdots < e_m $ and $R$ the matrix formed by  the associated linearly independant eigenvectors $r_1, \cdots , r_m $. Formulating the problem in the eigenspace, $W:=RX$, we obtain a decoupled system of scalar $m$ scalar hyperbolic problems:
\[\begin{array}{ll} \partial_t W^{(i)}(t,x) + \Lambda \partial_x W^{(i)}(t,x)  = 0, &\text{for } i=1,\cdots, m \end{array}.\]
The exact and numerical solutions of the such equations are well known in the literature, the reader can find a summary in \cite{Toro09, Vreugdenhil94}.

\begin{figure}[h!]
        \centering
        \includegraphics[width=0.8\textwidth]{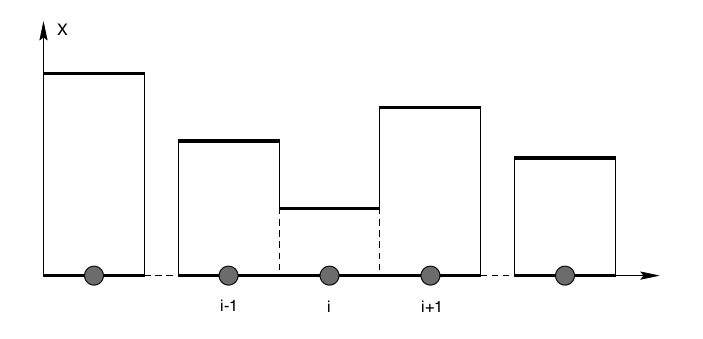}
        \caption{$X_i^n$ is built up from cell-averaged constant states}
        \label{piecewise}
\end{figure}

In the nonlinear case, we can resort to the Godunov method which assumes a piecewise constant
distribution of the data. Formally, this is carried out by defining cell averages:
\begin{equation}
	X_i^n = \frac{1}{\Delta x} \displaystyle \int_{x_{i-\frac{1}{2}}}^{x_{i+\frac{1}{2}}}X(x,t^n)dx ,
\end{equation}
which produce the desired piecewise constant distribution within each cell $I_i = \lbrace x_{i-\frac{1}{2}} ; x_{i+\frac{1}{2}} \rbrace$. This distribution is illustrated in figure \ref{piecewise} \cite{Toro09}, it appears that we have to deal with discontinuities at each cell interface. 
Thus, we have to solve a Riemann problem for each jump at a cell interface.
	
The Godunov method carries out to the exact solution of the Riemann problems at each cell interface. This approach has some advantages and disadvantages described in \cite{Vreugdenhil94}. As the construction of exact Riemann solutions is not a simple task, and we have to keep in mind the need to build its corresponding tangent and adjoint equations for the standard 4DVar implementation, we chose to rely on an approximate Riemann solver. Among several Riemann solvers we can find in the literature reviewed in \cite{Toro09}; we chose one of the most used approximate Riemann solver due to Roe \cite{Roe81} that we will briefly describe below.

In order to approximate the Riemann problem, Roe introduced the Jacobian matrix  $A = \frac{\partial F}{\partial X}$. By using the chain rule, the conservation laws \ref{conservativeNL} turn out to:
$$\partial_t X(t,x) + A \partial_x X(t,x)  = 0.$$
Roe's approach replaces the Jacobian matrix $A$ by a constant Jacobian matrix:
$$ \tilde A = \tilde A(X_L,X_R), $$
which is a function of the data states $X_L$ and $X_R$ on either sides of a given cell interface. Such matrix has to be cautiously chosen in order to satisfy the properties of hyperbolicity, consistency and conservation across discontinuities, which are detailed in \cite{Toro09}. Once we determine this matrix, we obtain a new linear system with constant coefficients that can be solved as previously.

Priestley \cite{Priestley87} and Glaister \cite{Glaister93} determined a suitable matrix for the SWE, we only give here the main results.
At each cell interface, the flux difference between the left  (L) and right (R) must be computed. In the eigenspace, this comes to:
$$\tilde A (X_R-X_L) = F(X_L) - F(X_R) = \displaystyle \sum_i a_i c_i r_i.$$
Introducing the auxiliary variable $w := (w_1\; w_2\; w_3)\transp = (\sqrt{h}  \; u\sqrt{h} \;v \sqrt{h})\transp$ the eigenvalues are defined by
$$\begin{array}{lll}
			c_1=\frac{\bar w_2}{\bar w_1} - \sqrt{w_1^2}, &
			c_2=\frac{\bar w_2}{\bar w_1} ,&
			c_3=\frac{\bar w_2}{\bar w_1} +- \sqrt{w_1^2}
\end{array},$$
and their corresponding eigenvectors are
$$\begin{array}{lll}
			r_1=\left ( \begin{array}{c} 
				\bar w_1 \\ 
				\bar w_2 - \bar w_1 \sqrt{w_1^2}\\ 
				\bar w_3
			\end{array} \right ) &
			r_1=\left ( \begin{array}{c} 
				0 \\ 
				0\\ 
				w_1
			\end{array} \right ) &
			r_1=\left ( \begin{array}{c} 
				\bar w_1 \\ 
				\bar w_2 + \bar w_1 \sqrt{w_1^2}\\ 
				\bar w_3
			\end{array} \right )			
\end{array},$$
with the coefficients
$$\begin{array}{lll}
			a_1=\Delta w_1 - \frac{\bar w_1 \Delta w_2 - \bar w_2 \Delta w_1}{2 \bar w_1 \sqrt{w_1^2}}, &
			a_2=\frac{\bar w_1 \Delta w_3 - \bar w_3 \Delta w_1}{\bar w_1 }	,&
			a_3=\Delta w_1 + \frac{\bar w_1 \Delta w_2 - \bar w_2 \Delta w_1}{2 \bar w_1 \sqrt{w_1^2}}
\end{array}.$$

\section*{References}
\bibliography{biblio}

\end{document}